\documentclass[onecolumn]{IEEEtran}
\usepackage{multicol}
\usepackage{times}
\usepackage{times,amssymb,amsmath,amsfonts,float,color,cite,bbm,mathrsfs,float,stmaryrd,amsbsy,mathbbol}
\usepackage{amsfonts}
\usepackage{latexsym}
\usepackage{theorem}
\usepackage{graphicx}
\usepackage{subcaption}
\usepackage{enumitem}
\usepackage{algorithm,algpseudocode}
\usepackage[normalem]{ulem}
\usepackage{xcolor}
\newtheorem{theorem}{Theorem}[section]
\newtheorem{lemma}[theorem]{Lemma}
\newtheorem{claim}[theorem]{Claim}

\newtheorem{corollary}[theorem]{Corollary}
\newtheorem{definition}[theorem]{Definition}

\newtheorem{example}[theorem]{Example}
\theoremstyle{definition}
\newtheorem{remark}[theorem]{Remark}

\renewcommand{\thefigure}{\thesection.\arabic{figure}}
\usepackage{amsmath}

\allowdisplaybreaks

\makeatletter
\renewcommand{\@endtheorem}{\endtrivlist}
\makeatother

\newcommand\remove[1]{}

\makeatletter
\renewcommand{\thefigure}{{\@arabic\c@figure}}
\renewcommand{\fnum@figure}{{\bf Figure\,\thefigure}}
\makeatother

\newcommand\nc\newcommand
\nc{\cA}{\mathcal{A}}\nc{\cB}{\mathcal{B}}\nc{\cC}{\mathcal{C}}\nc{\cD}{\mathcal{D}}
\nc{\cE}{\mathcal{E}}\nc{\cF}{\mathcal{F}}\nc{\cG}{\mathcal{G}}\nc{\cH}{\mathcal{H}}
\nc{\cI}{\mathcal{I}}\nc{\cJ}{\mathcal{J}}\nc{\cK}{\mathcal{K}}\nc{\cL}{\mathcal{L}}
\nc{\cM}{\mathcal{M}}\nc{\cN}{\mathcal{N}}\nc{\cO}{\mathcal{O}}\nc{\cP}{\mathcal{P}}
\nc{\cQ}{\mathcal{Q}}\nc{\cR}{\mathcal{R}}\nc{\cS}{\mathscr{S}}\nc{\cT}{\mathcal{T}}
\nc{\cU}{\mathcal{U}}\nc{\cV}{\mathcal{V}}\nc{\cW}{\mathcal{W}}\nc{\cX}{\mathcal{X}}
\nc{\cY}{\mathcal{Y}}\nc{\cZ}{\mathcal{Z}}

\nc{\bba}{\mathbf{a}}\nc{\bbb}{\mathbf{b}}\nc{\bbc}{\mathbf{c}}\nc{\bbd}{\mathbf{d}}
\nc{\bbe}{\mathbf{e}}\nc{\bbf}{\mathbf{f}}\nc{\bbg}{\mathbf{g}}\nc{\bbh}{\mathbf{h}}
\nc{\bbi}{\mathbf{i}}\nc{\bbj}{\mathbf{j}}\nc{\bbk}{\mathbf{k}}\nc{\bbl}{\mathbf{l}}
\nc{\bbm}{\mathbf{m}}\nc{\bbn}{\mathbf{n}}\nc{\bbo}{\mathbf{o}}\nc{\bbp}{\mathbf{p}}
\nc{\bbq}{\mathbf{q}}\nc{\bbr}{\mathbf{r}}\nc{\bbs}{\mathbf{s}}\nc{\bbt}{\mathbf{t}}
\nc{\bbu}{\mathbf{u}}\nc{\bbv}{\mathbf{v}}\nc{\bbw}{\mathbf{w}}\nc{\bbx}{\mathbf{x}}
\nc{\bby}{\mathbf{y}}\nc{\bbz}{\mathbf{z}}

\nc{\bbA}{\mathbf{A}}\nc{\bbB}{\mathbf{B}}\nc{\bbC}{\mathbf{C}}\nc{\bbD}{\mathbf{D}}
\nc{\bbE}{\mathbf{E}}\nc{\bbF}{\mathbf{F}}\nc{\bbG}{\mathbf{G}}\nc{\bbH}{\mathbf{H}}
\nc{\bbI}{\mathbf{I}}\nc{\bbJ}{\mathbf{J}}\nc{\bbK}{\mathbf{K}}\nc{\bbL}{\mathbf{L}}
\nc{\bbM}{\mathbf{M}}\nc{\bbN}{\mathbf{N}}\nc{\bbO}{\mathbf{O}}\nc{\bbP}{\mathbf{P}}
\nc{\bbQ}{\mathbf{Q}}\nc{\bbR}{\mathbf{R}}\nc{\bbS}{\mathbf{S}}\nc{\bbT}{\mathbf{T}}
\nc{\bbU}{\mathbf{U}}\nc{\bbV}{\mathbf{V}}\nc{\bbW}{\mathbf{W}}\nc{\bbX}{\mathbf{X}}
\nc{\bbY}{\mathbf{Y}}\nc{\bbZ}{\mathbf{Z}}

\nc{\bfA}{\mathbb{A}}\nc{\bfB}{\mathbb{B}}\nc{\bfC}{\mathbb{C}}\nc{\bfD}{\mathbb{D}}
\nc{\bfE}{\mathbb{E}}\nc{\bfF}{\mathbb{F}}\nc{\bfG}{\mathbb{G}}\nc{\bfH}{\mathbb{H}}
\nc{\bfI}{\mathbb{I}}\nc{\bfJ}{\mathbb{J}}\nc{\bfK}{\mathbb{K}}\nc{\bfL}{\mathbb{L}}
\nc{\bfM}{\mathbb{M}}\nc{\bfN}{\mathbb{N}}\nc{\bfO}{\mathbb{O}}\nc{\bfP}{\mathbb{P}}
\nc{\bfQ}{\mathbb{Q}}\nc{\bfR}{\mathbb{R}}\nc{\bfS}{\mathbb{S}}\nc{\bfT}{\mathbb{T}}
\nc{\bfU}{\mathbb{U}}\nc{\bfV}{\mathbb{V}}\nc{\bfW}{\mathbb{W}}\nc{\bfX}{\mathbb{X}}
\nc{\bfY}{\mathbb{Y}}\nc{\bfZ}{\mathbb{Z}}

\nc{\sA}{\mathsf{A}}\nc{\sB}{\mathsf{B}}\nc{\sC}{\mathsf{C}}\nc{\sD}{\mathsf{D}}
\nc{\sE}{\mathsf{E}}\nc{\sF}{\mathsf{F}}\nc{\sG}{\mathsf{G}}\nc{\sH}{\mathsf{H}}
\nc{\sI}{\mathsf{I}}\nc{\sJ}{\mathsf{J}}\nc{\sK}{\mathsf{K}}\nc{\sL}{\mathsf{L}}
\nc{\sM}{\mathsf{M}}\nc{\sN}{\mathsf{N}}\nc{\sO}{\mathsf{O}}\nc{\sP}{\mathsf{P}}
\nc{\sQ}{\mathsf{Q}}\nc{\sR}{\mathsf{R}}\nc{\sS}{\mathsf{S}}\nc{\sT}{\mathsf{T}}
\nc{\sU}{\mathsf{U}}\nc{\sV}{\mathsf{V}}\nc{\sW}{\mathsf{W}}\nc{\sX}{\mathsf{X}}
\nc{\sY}{\mathsf{Y}}\nc{\sZ}{\mathsf{Z}}

\nc{\bi}{\mathbb{i}}

\newcommand{\mathset}[1]{\left\{#1\right\}}
\newcommand{\abs}[1]{\left|#1\right|}

\newcommand{\floorenv}[1]{\left\lfloor #1 \right\rfloor}
\newcommand{\parenv}[1]{\left( #1 \right)}
\newcommand{\sparenv}[1]{\left[ #1 \right]}

\nc{\set}[1]{\llbracket #1 \rrbracket}
\nc{\bmat}[1]{\begin{bmatrix} #1 \end{bmatrix}}

\newcommand{\bal}[1]{\begin{align}\label{#1}}
\newcommand{\eal}{\end{align}}
\renewcommand{\leq}{\leqslant}

\renewcommand{\geq}{\geqslant}

\renewcommand{\Bbb}{\mathbb}

\newcommand{\Tref}[1]{Theo\-rem\,\ref{#1}}

\newcommand{\Cref}[1]{Co\-ro\-lla\-ry\,\ref{#1}}
\renewcommand{\Bbb}{\mathbb}

\newcommand{\F}{{\Bbb F}}

\newcommand{\C}{{\Bbb C}}
\newcommand{\N}{{\Bbb N}}

\newcommand{\R}{{\Bbb R}}
\newcommand{\Z}{{\Bbb Z}}
\newcommand{\E}{{\Bbb E}}

\newcommand{\cAS}{\cA^{\star}}

\newcommand{\fr}{\mathsf{fr}}
\newcommand{\ct}{\mathsf{ct}}
\newcommand{\bfr}{\mathsf{\mathbf{fr}}}
\newcommand{\bct}{\mathsf{\mathbf{ct}}}
\newcommand{\bbeta}{\boldsymbol{\beta}}
\newcommand{\tl}{\triangleleft}
\newcommand{\0}{\mathbf{0}}
\newcommand{\1}{\mathbf{1}}
\newcommand{\bo}{\mathbbm{1}}
\nc{\vt}{\vartheta}
\nc{\vtp}{\vartheta_{\bbP}}
\nc{\vtpk}{\vartheta_{\bbP,k}}

\DeclareMathOperator{\supp}{Supp}

\DeclareMathOperator{\var}{Var}

\outer\def\proclaim #1. #2\par{\medbreak
 \noindent{\bf#1.\enspace}{\sl#2\par}%
 \ifdim\lastskip<\medskipamount \removelastskip\penalty55\medskip\fi}

\begin{document}

% paper title
\title{{On the Long-Term Behavior of $k$-tuples Frequencies in Mutation Systems}}

\author{\IEEEauthorblockN{Ohad Elishco}% \and\hspace*{.5in} \IEEEauthorblockN{Alexander Barg} 
}

\maketitle

{\renewcommand{\thefootnote}{}\footnotetext{

\vspace{-.2in}
 
\noindent\rule{1.5in}{.4pt}

{ %This work was presented in part at the 2021 IEEE International Symposium on Information Theory.

Ohad Elishco is with the School of Electrical and Computer Engineering, Ben-Gurion University of the Negev, 
Beer Sheva, Israel. Email: elishco@gmail.com. 
This research was supported by the Israel Science Foundation (Grant No. 1789/23).
}
%\vspace{-.1in}
}}

%\maketitle

%%%%%%%%%%%%%%%%%%%%%%%%%%%%%
%%%%%%%%%%%%%%%%%%%%%%%%%%%%%
\begin{abstract}
In response to the evolving landscape of data storage, researchers have increasingly explored non-traditional platforms, with DNA-based storage emerging as a cutting-edge solution. Our work is motivated by the potential of in-vivo DNA storage, known for its capacity to store vast amounts of information efficiently and confidentially within an organism's native DNA. While promising, in-vivo DNA storage faces challenges, including susceptibility to errors introduced by mutations. 
To understand the long-term behavior of such mutation systems, we investigate the frequency of $k$-tuples after multiple mutation applications.

Drawing inspiration from related works, we generalize results from the study of mutation systems, particularly focusing on the frequency of $k$-tuples. 
In this work, we provide a broad analysis through the construction of a specialized matrix and the identification of its eigenvectors. In the context of substitution and duplication systems, we leverage previous results on almost sure convergence, equating the expected frequency to the limiting frequency. Moreover, we demonstrate convergence in probability under certain assumptions.

\end{abstract}
%%%%%%%%%%%%%%%%%%%%%%%%%%%%%
%%%%%%%%%%%%%%%%%%%%%%%%%%%%%

\section{Introduction}\label{Sec:Intro}
%%%%%%%%%%%%%%%%%%%%%%%%%%%%%%%%%%%%%%%
%%%%%%%%%%%%%%%%%%%%%%%%%%%%%%%%%%%%%%%

In recent years, the field of data storage has undergone a remarkable transformation, as researchers and practitioners have shifted their focus towards exploring alternative platforms that go beyond traditional methods. One such platform that has captured the attention of experts is DNA-based storage. By leveraging the inherent information density and stability of DNA molecules, this approach offers unparalleled advantages in terms of data density, longevity, and sustainability \cite{balado2012capacity,wong2003organic,yazdi2017portable}. 

This research is motivated by the emerging field of in-vivo DNA storage, which is characterized by its capacity to store vast amounts of information in a condensed and resilient format. 
Within the topic of information theory and coding, DNA-based storage systems have been investigated along two parallel trajectories. 
The first avenue studies in-vitro storage, in which information is synthesized and stored in a specialized container. 
In-vitro storage research is dedicated to improving the synthesis process \cite{lenz2020coding,elishco2023optimal,abu2023dna,makarychev2022batch,jain2020coding}, 
protecting against errors that occur in the synthesis process and during storage (due to stability issues) \cite{blawat2016forward,drinea2007improved,erlich2017dna,liu2022capacity,nguyen2021capacity,cai2021correcting,nguyen2020constrained,lenz2019coding,yin2021design}, 
and preventing errors that may occur in the reading process due to technical limitations of current techniques \cite{bhardwaj2021trace,magner2016fundamental,shafir2021sequence,elishco2021repeat,marcovich2021reconstruction}.

The second avenue explores in-vivo DNA storage, where information is stored within the DNA of a living organism. 
While in-vivo storage may not match the efficiency of in-vitro storage, it presents a unique advantage by seamlessly integrating stored information with an organism's native DNA, adding an extra layer of confidentiality \cite{clelland1999hiding}. 
Additionally, it holds potential for applications such as watermarking genetic modifications, facilitating reliable replication, and ensuring data protection \cite{heider2007dna,liss2012embedding,shipman2017crispr,jupiter2010dna}.

While in-vivo DNA storage presents a promising avenue for information storage, it is not without its challenges. 
One notable drawback is the inherent susceptibility to errors introduced by mutations. 
These mutations, including duplications, deletions, and substitutions, can compromise the stored information. 
Various mutation systems were studied, sometimes categorizing them under distinct names such as duplication systems or substitution systems, contingent on the specific type of mutation under investigation. Diverse aspects of mutation systems have been explored, including their properties \cite{jain2017capacity,ben2022reverse}, entropy \cite{lou2019evolution,elishco2019entropy,farnoud2015capacity}, and the development of error-correcting codes \cite{kovavcevic2018asymptotically,kovacevic2019codes,lenz2018bounds,lenz2017bounds,tang2021error,lenz2019duplication,jain2017duplication,mahdavifar2017asymptotically,sala2017exact,tang2021error,tang2020single}. 

The errors studied in both in-vivo and in-vitro storage systems overlap in some cases (e.g., edit errors occur in both storage types); however, the primary distinction lies in how the errors are introduced. 
In in-vitro storage systems, errors occur during the synthesis, storage, and reading processes of an information sequence with a known length, although deletion errors may shorten the length of the word. 
Conversely, in in-vivo storage, the information sequence evolves over time due to mutations that may increase its length, and drastically affect the composition of the sequence.

Our goal in this paper is to thoroughly investigate mutation systems, shedding light on their long-term behavior by examining the frequency of $k$-tuples following repeated mutation applications. 
Drawing inspiration from related works \cite{farnoud2019estimation,elishco2019entropy,ben2022reverse,lou2019evolution}, we seek to generalize and extend their results. 
Among these, \cite{lou2019evolution} is particularly relevant, wherein the authors explore the frequencies of $k$-tuples in two mutation systems.
The first system incorporates substitutions and duplications of variable lengths, while the second involves interspersed duplication. 
In both cases, the authors demonstrate the almost sure convergence of $k$-tuple frequencies and propose a mechanism to determine the limit. 
However, when attempting to extend this approach, a key limitation arises due to its reliance on solving a differential equation to determine the limit, which may prove challenging in certain scenarios.

In a sense, the analysis in \cite{lou2019evolution} laid the groundwork for this paper. 
Although the results in \cite{lou2019evolution} exhibit greater strength, they apply to a narrower range of mutation systems and necessitate solutions to vector differential equations. In contrast, our study encompasses a somewhat broader family of mutations, focusing mostly on the expected frequency of $k$-tuples and not on the frequencies. The determination of this expected frequency is obtained from the construction of a specialized matrix, and the identification of its eigenvectors associated with maximal eigenvalues. For substitution and duplication systems, the results in \cite{lou2019evolution}, which establish almost sure convergence, imply that for those mutation systems, the expected frequency of $k$-tuples is equivalent to the limiting frequency. 
It is worth noting that under certain assumptions, we also demonstrate the convergence in probability of the frequency of $k$-tuples and determine the limit using eigenvectors associated with maximal eigenvalues.  
As demonstrated in \cite{lou2019evolution}, the frequency of $k$-tuples proves valuable for bounding the entropy of a mutation system by defining a suitable semi-constrained system \cite{elishco2016semiconstrained}. Since the entropy function is concave, the expected frequency of $k$-tuples can also be utilized to provide an upper bound for the entropy of a mutation system.

The paper is organized as follows. Section \ref{Sec:Pre} contains notations, definitions, and useful approximations. It also compiles some established results on non-negative matrices, along with presenting new findings that will be employed, some of which may hold independent interest. 
In Section \ref{sec:l_mut}, our focus is directed toward the study of mutation systems characterized by a fixed mutation length. While this represents the most restrictive property, the outcomes in this section are notably robust, furnishing the expected frequency of $k$-tuples at any step, not solely at the limit. 
In Section \ref{sec:Avg_l_mut_sys}, we relax the constraint of a fixed mutation length and delve into the study of mutations with a fixed average length. 
Initially, we determine the limiting expected frequency of $k$-tuples. 
Subsequently, with additional assumptions, we ascertain the limit (in probability) of the frequency of $k$-tuples. 
Finally, Section \ref{sec:conc} offers concluding remarks for the paper.

\section{Preliminaries}\label{Sec:Pre}
%%%%%%%%%%%%%%%%%%%%%%%%%%%%%%%%%%%%%%
%%%%%%%%%%%%%%%%%%%%%%%%%%%%%%%%%%%%%%

Let $\cA=\mathset{a_0,\dots,a_{d-1}}$ be a finite alphabet of size $|\cA|=d$. 
A word (or a string) $w=w_0w_1\dots w_{n-1}$ over $\cA$ is a finite concatenation of symbols from $\cA$. 
The notation $|\cdot|$ serves a dual purpose, indicating both set size and word length.  
The set $\cA^k$ denotes all $k$-length words (also called $k$-tuples) over $\cA$, and $\cAS:=\bigcup_{m=1}^{\infty}\cA^m$ is the set of all finite words over $\cA$. 

For an integer $n\in\N$, we denote $[n]:=\mathset{0,1,\dots, n-1}$ and for $r\in \Z$, we define $r+[n]=\mathset{r+j ~:~ j\in [n]}$. 
In general, if $A\subseteq \Z$ and $r\in \Z$ we write $r+A:=\mathset{r+j ~:~ j\in A}$. 
For a word $w\in \cA^n$ and for $A\subseteq [n]$, we denote by $w_{[A]}$ the restriction of $w$ to the positions in $A$. 
For $0\leq i<j<n$, we also write $w_i^j$ to denote $w_i^j=w_{i+[j-i+1]}=w_i w_{i+1} \dots w_j$.

For $u,v\in\cAS$, $uv$ denotes the concatenation of $u$ and $v$, and $u^n$ represents the concatenation of $u$ with itself $n$ times. 
We say that $u$ is a subword of $v$, denoted $u\tl v$, if $|u|\leq |v|$ and there exists $i\in [|v|]$ such that $v_{i+[|u|]}=u$, 
with coordinates taken modulo $|v|$.

We denote by $\ct_v(u)$ the number of times $u$ appears in $v$ as a subword, i.e., 
\[\ct_v(u):=\sum_{i=0}^{|v|} \bo_{[u=v_{i+[|u|]}]}\]
where coordinates are taken modulo $|v|$ and 
\[\bo_{[a=b]}=\begin{cases} 1 & \text{ if } a=b \\ 0 & \text{ otherwise}\end{cases}\] 
is the indicator function. 
We denote by $\fr_v(u)$ the frequency of $u$ in $v$, i.e., 
\[\fr_v(u)=\begin{cases} 0 & \text{ if } |u|>|v| \\
\frac{1}{|v|}\ct_v(u) & \text{ if } |u|\leq |v|\end{cases}.\]

For a given $k\in\N$, the count vector $\bct^{(k)}_v\in\R^{d^{k}}$ has coordinates corresponding to $k$-tuples. For any $u\in\cA^{k}$, $\bct^{(k)}_v(u)$ represents the count of occurrences of $u$ in $v$. When $k$ is evident or does not impact the result, we simplify it to $\bct_v$. Similarly, $\bfr^{(k)}_v$ denotes the vector of frequencies.

We now define a mutation. 
\begin{definition} 
\label{def:mutation_law}
Let $\cA=\mathset{a_0,\dots,a_{d-1}}$ represent a finite alphabet with a size of $|\cA|=d$. 
A \textbf{mutation law} is a pair $(\vt,\bbP)$, where $\bbP=\mathset{\bbP_{a_t}}_{t\in [d]}$ is a set of $d$ non-degenerate, 
finitely supported probability measures on $\cAS$, and $\vt:\cA\to \cAS$ is a random function. 
For any $a_t\in\cA$, $\vt(a_t)$ is a random word chosen according to $\bbP_{a_t}$.

We denote the support of $\bbP_{a_t}$ as $\supp(\bbP_{a_t})=\mathset{w\in\cAS : \bbP_{a_t}(w)>0}$ and its cardinality as 
$|\supp(\bbP_{a_t})|=r_t$. 
If there exists $\tau\in\N$ such that for every $i\in [d]$, $\supp(\vt(a_t))\subseteq \cA^\tau$, we say that $(\vt,\bbP)$ is an 
\textbf{$\tau$-mutation law}.

In other words, $\bbP$ is a set of non-degenerate probability vectors 
\[\bbP:=\mathset{\bbp_{a_t}=(p_{a_t,0},p_{a_t,1},\dots,p_{a_t,r_t-1}) ~:~ \forall t\; \bbp_{a_t}\in [0,1]^{r_t},\; \sum_{j\in [r_t]}p_{a_t,j}=1},\]
such that for every $a_t\in \cA$, if $\supp(\vt(a_t))=\mathset{w_0,\dots,w_{r_t-1}}\subseteq \cAS$ then 
\[\vt(a_t)= \begin{cases} 
w_0 & \text{ with probability } p_{a_t,0}\\
\vdots \\
w_{r_t-1}& \text{ with probability } p_{a_t,r_t-1}
\end{cases}.\] 
\end{definition}
We stress that in our definition, a symbol cannot be deleted, but must be replaced with a word that contains at least one symbol. 
This implies that for every symbol $a_t\in\cA$, $|\vt(a_t)|\geq 1$ with probability $1$.

\begin{example}
\label{mut_ex} 
The following examples demonstrate the idea. 
\begin{enumerate}
    \item Let $\cA=\mathset{a_0,\dots,a_{d-1}}$ be an alphabet of size $|\cA|=d$ and let $\vt(a_t)=a_t a_t$ for every $a_t\in\cA$. 
    In this case, $\supp(\vt(a_t))\in \cA^2$ so the law is a $2$-mutation law, also known as tandem duplication \cite{farnoud2015capacity}.

    \item Similarly to the first example, we can define a "noisy" duplication rule. Let $\cA=\mathset{a_0,\dots,a_{d-1}}$ be an alphabet of size $|\cA|=d$, 
    let $\vt(a_i)=\mathset{a_t b ~:~ b\in \cA}$, and $\bbP=\mathset{\bbp_{a_t}=(p_{a_t,0},\dots,p_{a_t,d-1})\in [0,1]^d ~:~ t\in [d]}$ such that 
    \[\vt(a_t)= \begin{cases}
    a_t a_0 &\text{ w.p. } p_{a_t,0} \\ 
    a_t a_1 &\text{ w.p. } p_{a_t,1} \\
    \vdots & \vdots \\ 
    a_t a_{d-1} &\text{ w.p. } p_{a_t,d-1}.
    \end{cases}\]
\end{enumerate}
\end{example}

We now define a mutation step (or mutation round). 
\begin{definition}
Let $w=w_0 \dots w_{m-1}\in\cAS$ be a word over an alphabet $\cA$ with $|w|=m$. 
Let $(\vt,\bbP)$ be a mutation law over $\cA$. 
A \textbf{mutation step} (or mutation round) is performed by first selecting a position $i\in [m]$ uniformly at random, 
and then applying $\vt$ on $w_i$ to obtain the random word $w_0\dots w_{i-1} \vt(w_i) w_{i+1}\dots w_{n-1}$. 
With a slight abuse of notation, we denote the obtained (random) word by $\vt(w)$. 

In other words, a mutation step on $w$ is a random word $\vt(w)$ such that 
\[\vt(w):= \begin{cases}
    \vt(w_0) w_1 \dots w_{m-1} & \text{ w.p. } \frac{1}{m} \\ 
    \vdots & \vdots \\
    w_0\dots w_{m-2}\vt(w_{m-1}) &\text{ w.p. } \frac{1}{m}.
\end{cases}\] 
\end{definition} 

The next definition encapsulates the evolution of a word under repeated mutations. 
\begin{definition}
    A \textbf{mutation system} is defined by $S=\parenv{\cA,w,(\vt,\bbP)}$, where $\cA$ is a finite alphabet, 
    $S(0)=w\in \cAS$ is a non-empty starting word, and $(\vt,\bbP)$ is a mutation law defined over $\cA$. 
    We call $S$ an $\tau$-mutation system if $(\vt,\bbP)$ is an $\tau$-mutation law. 
    A mutation system evolves according to a discrete random model defined recursively. For $n\in \N$, set $S(n)=\vt(S(n-1))=\vt^n(w)$, 
    where $\vt^n(w)=\vt(\vt(\dots( \vt(w))\dots ))$ is the random word obtained by application of $n$ mutation steps, 
    starting with the word $w$. 
\end{definition}

\begin{example}
Let $\cA=\mathset{\sA,\sG,\sT,\sC}$ and consider the noisy duplication rule from Example \ref{mut_ex}. 
Let $w=\mathsf{AGTC}\in\cA^4$ and assume at first that $i=1$ was chosen. 
Then $S(1)$, obtained after a single mutation step, is $\mathsf{AGATC}, \mathsf{AGGTC},\mathsf{AGTTC},\mathsf{AGCTC}$ with probabilities $\frac{p_{G,0}}{4},\frac{p_{G,1}}{4},\frac{p_{G,2}}{4},\frac{p_{G,3}}{4}$, respectively. 
\end{example} 

\begin{remark} 
    In this paper, a mutation law may transform a single symbol and not several symbols at once. 
    nevertheless, the results we obtain here can be generalized by considering sliding windows of length $k$ as symbols. 
    For instance, for a word $w=w_0 w_1\dots w_{n-1}$, define 
    \[w^{[k]}=\bmat{w_0\\ \vdots\\ w_{k-1}} \bmat{w_2\\ \vdots\\ w_k}\dots \bmat{w_{n-1}\\ \vdots\\ w_{k-2}}\] 
    where coordinates are taken modulo $n$. 
    A mutation law $\vt^{(k)}$ can be defined over $\cA^k$ by treating each $\bmat{w_i\\ \vdots\\ w_{i+k-1}}$ as a symbol. 
    Notice, however, that the set of obtained mutations does not encompass the entire set of mutation laws that can be defined on $k$-tuples. 
    Indeed, if we denote by $w^{[k]}$ the operation of treating each $k$-tuple in $w$ as a symbol, the inverse operation is obtained by reading the first symbol that appears in every $k$-tuple. 
    The set of obtained mutations is the set of all mutations for which $\parenv{\vt(w)}^{[k]}=\vt^{(k)}\parenv{w^{[k]}}$ for every word $w$.
    A family of mutations that can be defined by considering sliding block windows is the family of substitution and duplication systems, in which a (limited length) block is duplicated in each mutation step.
\end{remark} 

The gamma function $\Gamma(n+1)=n\Gamma(n)$ will appear numerous times in the paper. 
Almost every time, we apply Stirling's approximation for the Gamma function \cite[Eq. 25.15]{spiegel1990mathematical} which 
states that for any $0\neq x\in \C$, $\Gamma(x)=\sqrt{\frac{2\pi}{x}}\parenv{\frac{x}{e}}^x\parenv{1+O(1/x)}$. 
Thus, for large $n$, we can use Stirling's approximation to get 
\begin{align}
\label{eq:gamma_simple_approx}
\Gamma(n+x)\approx \Gamma(n)n^x.
\end{align}
Moreover, for $x,y\in\C$, consider products of the form 
\begin{align*}
\prod_{t=j}^n \frac{\parenv{t+x}}{t+y}&= \frac{\Gamma(n+1+x)}{\Gamma(j+1+x)}\frac{\Gamma(j+1+y)}{\Gamma(n+1+y)}. 
\end{align*}
Using the approximation, we obtain  
\begin{align*}
\frac{\Gamma(n+1+x)}{\Gamma(n+1+y)}&= \frac{\sqrt{\frac{2\pi}{n+1+x}}\parenv{\frac{n+1+x}{e}}^{n+1+x}\parenv{1+O(1/(n+1+k))}}{\sqrt{\frac{2\pi}{n+1+x}}\parenv{\frac{n+1+y}{e}}^{n+1+y}\parenv{1+O(1/(n+1+k))}}\\ 
&\sim n^{x-y}+O\parenv{n^{x-y-1}}. 
\end{align*}
Which, in turn, implies 
\begin{align}
    \label{eq:stirling_for_gamma_ratio}
    \prod_{t=j}^n \frac{t+x}{t+y}&\sim \parenv{\frac{n}{j}}^{x-y}+O\parenv{n^{x-y-1}}.
\end{align}

Throughout the paper, we will leverage various properties associated with non-negative matrices. 
For a field $\F$, and given $n, m \in \N$, we use $\cM^{n\times m}(\F)$ to represent the set of $n\times m$ matrices over $\F$. 
In most instances, we consider $\F$ to be the set of real numbers. 
If $\F$ is evident from the context, we will omit it and simply write $\cM^{n\times m}$.

For matrices $\bbM,\bbN\in \cM^{n\times m}(\R)$, the notation $\bbM\geq \bbN\; (\bbM>\bbN)$ implies that for every $i\in [n], j\in [m]$, 
$\bbM_{i,j}\geq \bbN_{i,j}\; (\bbM_{i,j}>\bbN_{i,j})$. 
Here, $\1^{n\times m}$ and $\0^{n\times m}$ denote the $n\times m$ all-ones matrix and the all-zero matrix, respectively. 
We will use $\1$ and $\0$ also to denote the all-one and the all-zero vectors. 
Additionally, $\bbI^{n,m}$ represents the $n\times m$ identity matrix. 
When the dimensions are clear from the context, we omit them from the notation and write $\1$, $\0$, and $\bbI$. 

A vector $\bbv$ is called a generalized eigenvector of $\bbM$, associated with the eigenvalue $\lambda$, 
if there is an integer $k\in\N$, such that $(\bbM-\lambda\bbI)^j\cdot \bbv\neq 0$ for $j<k$ and $(\bbM-\lambda\bbI)^k\cdot \bbv= 0$. 
This implies that there is a chain $\bbv_0,\bbv_1,\dots,\bbv_{k-1}=\bbv$ such that for $j\in [k]$, $\bbM\cdot \bbv_j=\lambda\bbv_j+\bbv_{j-1}$ where $\bbM\cdot\bbv_0=\lambda\bbv_0$. 
Every matrix $\bbM\in\cM^{n\times n}$ can be written as $\bbM=\bbA\Lambda\bbA^{-1}$ where $\Lambda$ is in Jordan canonical form, 
the rows of $\bbA^{-1}$ are the left eigenvectors (and left generalized eigenvectors) of $\bbM$, and the columns of $\bbA$ are the right eigenvectors and generalized eigenvectors of $\bbM$. 
The matrix $\Lambda$ is a block diagonal matrix in which every block $\Lambda_i$ is a Jordan block corresponding to an eigenvalue $\lambda_i$. 
The entries on the main diagonal of $\Lambda_i$ are $\lambda_i$ and the entries on the diagonal below the main diagonal are $1$. 
\[\Lambda_j=\begin{bmatrix} \lambda_j& & & &  \\1& \lambda_j & & & \\ & 1 &\lambda_j & & \\ & & \ddots &\ddots & \\ & & & 1& \lambda_j\end{bmatrix}.\]

A matrix $\bbM\in \cM^{n\times m}$ is called \textbf{non-negative} if $\bbM\geq \0$.
We use $\rho(\bbM)$ to denote the spectral radius of $\bbM$, which is defined as the maximum absolute value (modulus) of its eigenvalues 
\[\rho(\bbM)=\max \mathset{|\lambda| ~:~ \lambda \text{ is an eigenvalue of } \bbM}.\]

The following theorem states several known properties of non-negative matrices. 
\begin{theorem}\cite[Section 8.4]{horn2012matrix}
\label{th:non_neg_mat_properties} 
Let $\bbM\in\cM^{n\times n}$ be a non-negative square matrix. Then 
\begin{enumerate}
    \item $\rho(\bbM)$ is an eigenvalue of $\bbM$ and there is a non-negative, non-zero vector $\bbv$ such that $\bbM\cdot \bbv=\rho(\bbM)\bbv$. 
    \item If $\bbv$ is a non-negative, non-zero vector for which $\bbM\cdot\bbv\geq \alpha\bbv$ for some $\alpha\in\R$, then $\rho(\bbM)\geq \alpha$. 
    \item Suppose that there is a \underline{positive} vector $\bbv$ and a non-negative number $\alpha$ such that $\bbv\cdot \bbM=\alpha\bbv$. Then $\alpha=\rho(\bbM)$. 
    \item Suppose that $\bbM$ has a positive left eigenvector. If $\bbv$ is non-zero and $\bbM\cdot\bbv\geq \rho(\bbM)\bbv$ then $\bbv$ is an eigenvector of $\bbM$ corresponding to the eigenvalue $\rho(\bbM)$. 
    \item Suppose that $\bbM$ has a positive left eigenvector. If $\bbM\neq \0$ then $\rho(\bbM)>0$ and every eigenvalue $\lambda$ of $\bbM$ with maximum modulus $|\lambda|=\rho(\bbM)$ is semi-simple (algebraic multiplicity is equal to geometric multiplicity), that is, every Jordan block of $\bbM$ that corresponds to a maximum modulus eigenvalue is one-by-one.
\end{enumerate}  
\end{theorem}

A non-negative matrix $\bbM\in\cM^{n\times n}$ is called \textbf{irreducible} iff $(\bbI+\bbM)^{n-1}>0$. 
The following result follows immediately from the definition. 
\begin{corollary}
    \label{cor:sum_of_mat_is_irreducible} 
    Let $\bbM,\bbN\in\cM^{n\times n}$ be non-negative matrices and assume $\bbM$ is irreducible. Then $\bbM+\bbN$ is irreducible. 
\end{corollary}
Theorem \ref{th:non_neg_mat_properties} can be strengthened if $\bbM$ is an irreducible matrix. 
\begin{theorem}\cite[Section 8.4]{horn2012matrix}
\label{th:irreducible_mat_properties} 
Let $\bbM\in\cM^{n\times n}$ be a non-negative, irreducible square matrix. Then 
\begin{enumerate}
    \item $\rho(\bbM)>0$ is a simple eigenvalue of $\bbM$. 
    \item The left and right eigenvectors associated with $\rho(\bbM)$ are the only strictly positive eigenvectors. 
    \item The maximum modulus eigenvalues of $\bbM$ are $e^{2\pi\bbi p/k}\rho(\bbM)$ for $p\in [k]$, and each has an algebraic multiplicity $1$. 
    \item if $\bbM\geq \bbN$ are real, non-negative matrices and $\bbM$ is irreducible, then $\rho(\bbM)\geq \rho(\bbN)$, with equality iff $\bbM=\bbN$. 
\end{enumerate}
\end{theorem}

%\begin{remark}
%    Some of the properties that are stated in Theorem \ref{th:irreducible_mat_properties} are a part of the well-known Perron-Frobenius Theorem. 
%\end{remark}

When a non-negative matrix $\bbM$ has the additional property of a constant column sum (also known as stochastic matrices), additional properties can be found. 
Before presenting the additional properties, notice that if the entries of every column in $\bbM$ sum to $\alpha$, then the all-ones vector $\1$ is a left eigenvector associated with the eigenvalue $\alpha$, $\1\cdot\bbM=\alpha\1$. 
Thus, Theorem \ref{th:non_neg_mat_properties} implies that $\rho(\bbM)=\alpha$, and that there are no generalized eigenvectors associated with maximum modulus eigenvalues. 
In addition, without loss of generality, we may assume that $\bbM$ is a (lower) block-triangular matrix, 
such that every block is \underline{irreducible} when considered as a matrix (see, for example, \cite[Section 4.4]{LinMar21}). 
This means, that there are $t$ irreducible square matrices $\bbB_0,\bbB_1,\dots, \bbB_{t-1}$, with dimensions $n_0\times n_0,\dots, n_{t-1}\times n_{t-1}$, 
respectively, such that 
\[\bbM=\bmat{\bbB_0&\0&\0&\dots &\0\\ *_{1,0}&\bbB_1&\0&\dots&\0 \\ *_{2,0}&*_{2,1}&\bbB_2&\dots&\0 \\ \vdots&\vdots&\ddots&\dots&\vdots\\ *_{t-1,0}&*_{t-1,1}&*_{t-1,2}&\dots&\bbB_{t-1}},\] 
where $*_{i,j},\; i>j$ represents blocks of dimension $n_i\times n_j$ that may contain positive values. 
The eigenvalues of $\bbM$ are the eigenvalues of the sub-matrices $\bbB_j$. 
Since each of the sub-matrices $\bbB_j$ is irreducible, it has a unique real maximal eigenvalue. 
This implies that the number of sub-matrices $\bbB_j$ associated with $\alpha$ is equal to the algebraic multiplicity of $\alpha$ in the characteristic polynomial of $\bbM$. 
In addition, every real maximal eigenvalue $\rho(\bbM)=\alpha$ is associated with its own unique sub-block $\bbB_j$ and every maximum modulus eigenvalue (not necessarily real) is associated with a sub-matrix $\bbB_j$ with eigenvalue $\alpha$ (there could be a block $\bbB_j$ with several eigenvalues of maximum modulus, only one of them is real). 
Our first goal is to show that \underline{right} eigenvectors associated with eigenvalues of maximum modulus have pairwise disjoint support. 
To that end, we define the support of an $n$-dimensional vector $\bbv$ as the set of coordinates that contain a non-zero value 
\[\supp(\bbv)=\mathset{j\in [n] ~:~ \bbv_j\neq 0}.\] 

\begin{lemma}
\label{lem:right_eigenvec_disjoint_support}
    Let $\bbM\in\cM^{n\times n}$ be a non-negative square matrix in which the entries in every column sum to a constant $\alpha$. 
    In other words, if $\1$ is the $n$ dimensional row vector of all ones, then $\1\cdot\bbM=\alpha\1$. 
    Assume there are $s$ right vectors $\bbr_0,\dots \bbr_{s-1}$ associated with the eigenvalues $\rho(\bbM)=\alpha$. 
    Then for every $i,j\in [s]$, if $i\neq j$ then 
    \[\supp(\bbr_i)\cap\supp(\bbr_j)=\emptyset.\] 
\end{lemma}

\begin{IEEEproof} 
    First, notice that $\rho(\bbM)=\alpha$.  
    Assume that $\bbM$ is in a (lower) block-triangular matrix form 
    \[\bbM=\bmat{\bbB_0&\0&\0&\dots &\0\\ *_{1,0}&\bbB_1&\0&\dots&\0 \\ *_{2,0}&*_{2,1}&\bbB_2&\dots&\0 \\ \vdots&\vdots&\ddots&\dots&\vdots\\ *_{t-1,0}&*_{t-1,1}&*_{t-1,2}&\dots&\bbB_{t-1}}.\] 
    To prove the lemma, it suffices to show that if $\bbB_j$ has an eigenvalue $\alpha$, then for $i>j$, $*_{i,j}=\0$. 
    If $\bbv$ is a right eigenvector of $\bbB_j$, associated with $\alpha$, then padding $\bbv$ with $\sum_{i=0}^{j-1}n_i$ zeros from above and $\sum_{i=j+1}^{t-1}n_i$ zero from below yields a right eigenvector of $\bbM$ associated with $\alpha$. This, together with the property that the number of sub-matrices $\bbB_j$ with eigenvalue $\alpha$ is equal to the algebraic multiplicity of $\alpha$, finishes the proof. 
    
    To show that if $\bbB_j$ has an eigenvalue $\alpha$, then for $i>j$, $*_{i,j}=\0$, we generate an irreducible matrix $\bbB'_j$ by adding to $\bbB_j$ the values that appear in $*_{i,j},\; i>j$. We then claim that $\rho(\bbB'_j)=\alpha$ and $\bbB'_j\geq \bbB_j$. Theorem \ref{th:irreducible_mat_properties} implies that $\bbB'_j=\bbB_j$ which, in turn, implies that $*_{i,j}=0$ since the sum of each column in $\bbM$ is $\alpha$. 
    
    First, notice that the sub-matrix $\bbB_{t-1}$ must have an eigenvalue $\alpha$, as a square matrix in which the entries of every column sum to $\alpha$. 
    Fix $j\neq t-1$ for which $\bbB_j$ has an eigenvalue $\alpha$. 
    Generate a matrix $\bbB_j'$ as follows. 
    Let $\bmat{\bbI^{n_j\times n_j}& \bbA}$ be the $n_j\times \sum_{i=j}^{t-1} n_i$ matrix comprised of the identity matrix $\bbI^{n_j\times n_j}$ and an $n_j\times \sum_{i=j+1}^{t-1} n_i$ matrix $\bbA$ in which all the entries are zeros except for the last row in which all the entries are $1$,  
    \[\bbA=\bmat{0&0&\dots&0\\ \vdots&\vdots&\ddots&\vdots\\ 0&0&\dots&0\\ 1&1&\dots&1}.\] 
    Take the sub-matrix $\bmat{\bbB_j\\ *_{j+1,j}\\ \dots \\*_{t-1,j}}$ of dimensions $\parenv{\sum_{i=j}^{t-1} n_i}\times n_j$ and multiply it by $\bmat{\bbI^{n_j\times n_j}& \bbA}$ to obtain the $n_j\times n_j$ matrix 
    \[\bbB_j'=\bmat{ \bbI^{n_j\times n_j}&\bbA}\cdot \bmat{\bbB_j\\ *_{j+1,j}\\ \vdots \\ *_{t-1,j}}.\] 
    In other words, we add to the bottom entry of every column in $\bbB_j$, all the values that appear below it in $\bbM$. 
    According to Corollary \ref{cor:sum_of_mat_is_irreducible}, $\bbB_j'$ is irreducible, and it is clear that $\bbB_j'\geq \bbB_j$. 
    Since the sum of the entries of each column in $\bbM$ is $\alpha$, the sum of the entries of each column in $\bbB_j'$ is also $\alpha$, obtained with the left eigenvector $\1$ of all ones. By Theorem \ref{th:irreducible_mat_properties}, $\bbB_j'=\bbB_j$, which implies that for all $i>j$, $*_{i,j}=\0$. 
\end{IEEEproof}

Using similar arguments as in the proof of the previous lemma, we can study the support of left eigenvectors associated with maximal modulus eigenvalues. 
For simplicity of notation, we denote by $S_j=\parenv{\sum_{i=0}^{j-1}n_i} +[n_j]$ the set of row (and column) coordinates associated with the block $\bbB_j$. 

\begin{lemma}
    \label{lem:support_of_left_eigenvec}
    Let $\bbM\in\cM^{n\times n}$ be a non-negative square matrix in which the entries in every column sum to a constant $\alpha$, and assume $\bbM$ is in a lower triangular block matrix form.  
    Assume that $\bbB_j$ is a block associated with an eigenvalue $\rho(\bbM)=\alpha$ and let $\bbl_j$ be a left eigenvector with $\supp(\bbl_j)\cap S_j\neq \emptyset$. 
    Then the set $\mathset{\bbl_j}_j$ can be chosen such that  
    \[\supp(\bbl_j)\subseteq S_j\cup \bigcup_{\substack{i\in [j]\\ i\rightsquigarrow j}}S_i\] 
    where $i\rightsquigarrow j$ means that there is a sequence $i=i_0,i_1,\dots,i_m=j$ such that for every $t\in [m]$, $*_{i_{t+1},i_t}\neq \0$.
\end{lemma}
Lemma \ref{lem:support_of_left_eigenvec} is best demonstrated by an example. 
\begin{example}
\label{ex:exmpl_for_left_eigenvectors}
    Let $\bbM\in \cM^{5\times 5}$ be the matrix 
    \[\bbM=\bmat{1&0&0&0&0&0\\ 1&5&0&0&0&0\\ 1&0&2&0&0&0\\ 1&0&2&2&1&0\\ 1&0&0&3&4&0\\ 0&0&1&0&0&5},\] 
    so the sum of entries in each column is $\rho(\bbM)=\alpha=5$, the number of sub-blocks is $t=5$, and 
    \begin{align*}
    \bbB_0&=\bmat{1},\;\bbB_1=\bmat{5},\; \bbB_2=\bmat{2},\\ 
    \bbB_3&= \bmat{2&1\\3&4},\; \bbB_4=\bmat{5}.
    \end{align*} 
    The right eigenvectors associated with the eigenvalue $\alpha$ are 
    \[\bbr_1=\bmat{0\\ 1\\ 0\\ 0\\ 0\\ 0}, \bbr_3=\bmat{0\\ 0\\ 0\\ 1\\ 3\\ 0}, \bbr_4=\bmat{0\\ 0\\ 0\\ 0\\ 0\\ 1}.\] 
    As stated in Lemma \ref{lem:right_eigenvec_disjoint_support}, the support of the right eigenvectors is pairwise disjoint. 
    
    The left eigenvectors associated with the eigenvalue $\alpha$ are expected, by Theorem \ref{th:non_neg_mat_properties}, to be non-negative. 
    We find 
    \begin{align*}
    \bbl_1&=\bmat{\frac{1}{4}&1&0&0&0&0},\\ 
    \bbl_3&=\bmat{\frac{2}{3}&0&\frac{2}{3}&1&1&0}\\ 
    \bbl_4&=\bmat{\frac{1}{12}&0&\frac{1}{3}&0&0&1}.
    \end{align*} 
    Notice that
    \begin{align*}
    S_0&=\mathset{0},\; S_1=\mathset{1},\; S_2=\mathset{2}\\ 
    S_3&=\mathset{3,4},\; S_4=\mathset{5}.
    \end{align*}
    As in Lemma \ref{lem:support_of_left_eigenvec}: 
    \begin{enumerate}
        \item $\supp(\bbl_1)=S_1\cup S_0=\mathset{0,1}$ since $*_{1,0}=\bmat{1}\neq \0$. 
        \item $\supp(\bbl_3)=S_3\cup S_2\cup S_0=\mathset{0,2,3,4}$ since $*_{3,0}=\bmat{1\\1}\neq \0$ and $*_{3,2}=\bmat{2\\0}\neq \0$. 
        \item $\supp(\bbl_4)=S_4\cup S_2\cup S_0=\mathset{0,2,5}$ since $*_{5,2}=\bmat{1}\neq \0$. 
        Here, $S_0$ appears in the union, although $*_{4,0}=\0$, since $0\rightsquigarrow 4$ using the sequence $(0,2,4):\;*_{2,0}=\bmat{1}\neq \0$ and $*_{4,2}=\bmat{1}\neq\0$.
    \end{enumerate} 
    Moreover, $\bbl_1$ is the only left eigenvector associated with $\alpha$ that is supported on $S_1$, and similarly, 
    $\bbl_3\; (\bbl_4)$ is the only left eigenvector associated with $\alpha$ that is supported on $S_3\; (S_4)$. 
\end{example}

\begin{IEEEproof}[Proof of Lemma \ref{lem:support_of_left_eigenvec}] 
    Let $\bbB_j$ be a block associated with a maximal eigenvalue. 
    Assume that $i<j$ is the maximal value for which $\bbB_i$ is not associated with an eigenvalue $\alpha$. 
    We claim that there is a left eigenvector $\bmat{\bbu& \1^{1\times n_j}}$ where $\bbu$ has dimensions $1\times n_i$, such that 
    \begin{align}
    \label{eq:mat_ex}
    \bmat{\bbu& \1^{1\times n_j}}\cdot \bmat{\bbB_i& \0 \\ *_{i,j}& \bbB_j}=\alpha \bmat{\bbu& \1^{1\times n_j}}.
    \end{align}

    The vector $\1^{1\times n_j}$ is an eigenvector of $\bbB_j$, associated with $\alpha$. 
    Thus, we are left to show that there exists $\bbu$ such that $\bmat{\bbu& \1^{1\times n_j}}\cdot \bmat{\bbB_i\\ *_{i,j}}=\alpha \bbu$. 
    Consider the $n_j\times n_i$ matrix $*_{i,j}$. 
    Let $\1^{1\times n_j}$ be the row vector of all ones and notice that to show \eqref{eq:mat_ex} it suffices to find an $n_i$-dimensional row vector $\bbu$ such that 
    $\bbu\cdot\parenv{\bbB_i-\alpha \bbI^{n_i\times n_i}}=-\1^{1\times n_j}\cdot *_{i,j}$. 
    We may obtain $\bbu$ easily if the matrix $\bbB_i-\alpha \bbI^{n_i\times n_i}$ is invertible. 
    To see that it is, notice that the modulus of eigenvalues of $\bbB_i$ are strictly smaller than $\alpha$ (due to $\bbB_i$ being irreducible). 
    Therefore, $\bbB_i-\alpha\bbI^{n_i\times n_i}$ does not have a $0$ eigenvalue, which implies it is invertible. 

    After finding $\bbu$, we consider the maximal $i_1<i$ where $\bbB_{i_1}$ is not associated with eigenvalue $\alpha$, and with $*_{i,i_1}\neq \0$ or $*_{j,i_1}\neq \0$, and repeat the same arguments to find a $1\times n_{i_1}$ row vector $\bbu_1$ such that 
    \[\bmat{\bbu_1& \bbu& \1^{1\times n_j}}\cdot \bmat{\bbB_{i_0}&\0&\0 \\ *_{i,i_0}& \bbB_i &\0 \\ *_{j,i_0}& *_{j,i} & \bbB_j}=\alpha \bmat{\bbu_1& \bbu& \1^{1\times n_j}}.\]
    We continue the same way to construct a left eigenvector $\bbl_j$ for $\bbM$, associated with the eigenvalue $\alpha$. 
    According to the construction, the only coordinates that might be positive in $\bbl_j$ are those that associated with $S_i$ such that $i\rightsquigarrow j$. 
\end{IEEEproof}

The following corollary follows from Lemma \ref{lem:support_of_left_eigenvec} immediately.
\begin{corollary}
\label{cor:sum_left_eigenvectors_is_1}
    Let $\bbM\in\cM^{n\times n}$ be a non-negative square matrix in which the entries in every column sum to a constant $\alpha$. 
    Assume that the algebraic multiplicity of the eigenvalue $\rho(\bbM)=\alpha$ is $s$, and let $\bbl_{j_t},\; t\in [s]$ denote the left eigenvectors associated with $\alpha$, 
    constructed in the proof of Lemma \ref{lem:support_of_left_eigenvec}. 
    Normalize $\bbl_{j_t}$ such that it contains $1$ in the coordinates $S_{j_t}$. 
    Then 
    \[\sum_{t\in [s]}\bbl_{j_t}=\1,\] 
    where $\1$ is the all-ones row vector.
\end{corollary}

\begin{IEEEproof} 
    Denote $\bbv=\sum_{j\in [t]}\bbl_j$ and notice that from the construction in the proof of Lemma \ref{lem:support_of_left_eigenvec}, 
    $\bbv$ contains $1$ in coordinates $\cup_{t\in [s]}S_{j_t}$ that correspond to blocks that are associated with $\alpha$. 
    It is clear that $\bbv\cdot \bbM=\alpha \bbv$. 
    Now let $\1$ be the row vector of all-ones, which is an eigenvector of $\bbM$ associated with the maximal eigenvalue $\alpha$. 
    Thus, $(\bbv-\1)$ is an eigenvector of $\bbM$ associated with $\alpha$, that contains $0$ in the coordinates $\cup_{t\in [s]}S_{j_t}$.
    But this implies that $(\bbv-\1)$ is an eigenvector of the matrix $\bbM'$, obtained by eliminating (placing zeros in) the blocks $\bbB_{j_t}$ for $t\in [s]$. 
    Since $\bbM'$ does not contain any block that is associated with $\alpha$, the spectrum of $\bbM'$ is strictly less than $\alpha$, 
    Which, in turn, implies that $\bbv-\1=\0$.
\end{IEEEproof}

\begin{example}
    As a simple example, consider the matrix $\bbM$ from Example \ref{ex:exmpl_for_left_eigenvectors} with its left eigenvectors. 
    The eigenvectors $\bbl_1,\bbl_3,\bbl_4$ contain $1$ in $S_1,S_3,S_4$, respectively. 
    Summing the eigenvectors yields 
    \[\bbl_1+\bbl_3+\bbl_4=\bmat{1&1&1&1&1&1}.\]
\end{example}
In the next section, some of the properties we mentioned above are being used to study the frequency of $k$-tuples in $\tau$-mutation systems.

\section{$\tau$-mutation systems}\label{sec:l_mut}
%%%%%%%%%%%%%%%%%%%%%%%%%%%%%%%%%%%%%%%%%%%%%%%%%
In this section, we direct our attention towards the (relatively) straightforward scenario of $\tau$-mutation systems $S$, where each mutation has a fixed length, denoted as $\tau$. 
As we undergo each mutation step, a symbol is \underline{replaced} with a sequence of length $\tau$, resulting in an increase of the entire word's length by $\tau-1$. 
Our objective is to formulate a matrix that incorporates key information about the frequency of $k$-tuples in $S(n)$, enabling us to apply algebraic tools for studying the evolution of $S$ over time. 
To achieve this, we draw inspiration from urn models (interested readers are directed to \cite{mahmoud2008polya} and the many references provided therein). 
To illustrate the fundamental concept, we initiate our exploration by examining the frequency of symbols. Here, we introduce the substitution matrix, a pivotal tool for the analysis of mutation systems. 

\begin{definition}
\label{def:sub_mat}
    Let $(\vt,\bbP)$ be a mutation law over an alphabet $\cA=\mathset{a_0,\dots, a_{d-1}}$ of size $|\cA|=d$. 
    The \textbf{substitution matrix} $\bbM:=\bbM_{(\vt,\bbP)}$ is a real, $d\times d$ matrix, with non-negative entries, defined by 
    \[\bbM_{i,j}=\E\sparenv{\ct_{\vt(a_j)}\parenv{a_i}}=\sum_{\eta\in \cAS}\Pr(\vt(a_j)=\eta)\ct_{\eta}(a_i).\]
\end{definition} 

\begin{example}
    \label{ex:running}
     Let $\cA=\mathset{0,1}$ be the binary alphabet. 
     For $a\in \cA$ consider the duplication law 
    \[\vt(0)=\begin{cases} 
    00 & \text{ w.p. } 2/3 \\ 
    01 & \text{ w.p. } 1/3, 
    \end{cases},\qquad 
    \vt(1)=\begin{cases} 
    11 & \text{ w.p. } 3/4 \\ 
    00 & \text{ w.p. } 1/4. 
    \end{cases}\] 
     We calculate the top left entry (the $(0,0)$ coordinate) in $\bbM$. 
     To that end, we calculate the expected number of zeros in $\vt(0)$. 
    Notice that $\vt(0)=00$ with probability $2/3$ and $\vt(0)=01$ with probability $1/3$. 
    Thus, the expected number of zeros is given by $2\cdot 2/3+ 1\cdot 1/3=5/3$. 
    Continuing the same way, the substitution matrix is given by  
    \[\bbM=\bmat{5/3 & 1/2 \\ 1/3 & 3/2}.\]  
\end{example} 

The following theorem shows the relation between $\bbM$ and and evolution of mutation systems. 
\begin{theorem}
\label{th:basic1}
    Let $S$ be an $\tau$-mutation system over $\cA=\mathset{a_0,\dots,a_{d-1}}$ with a starting word $w\in\cAS$ of length $|w|=m$, and let $n\in\N$ be a positive integer. 
    Then the expected frequency of symbols in $S(n)=\vt^n(w)$ is given by 
    \[\E\sparenv{\bfr_{S(n)}}=\frac{\E\sparenv{\bct_{S(n)}}}{m+n(\tau-1)} = \frac{1}{m+n(\tau-1)}\prod_{j=0}^{n-1}\frac{1}{m+j(\tau-1)}(\bbM+(m+j(\tau-1)-1)\bbI)\cdot \bct_{w}.\]
\end{theorem} 

Before proving \Tref{th:basic1} we show a simple example.
\begin{example}
     Consider Example \ref{ex:running} with $w=00111$ (so $|w|=m=5$).  
    Recall that  
    \[\bbM=\bmat{5/3 & 1/2 \\ 1/3 & 3/2}.\] 
    The substitution matrix $\bbM$ can be written as 
    \[\bbM=\bmat{-3/5&3/5\\ 3/5 &2/5}\bmat{7/6 &0 \\ 0 &2}\bmat{-2/3&1\\1&1}\] 
    where 
    \[\bmat{-3/5&3/5\\ 3/5 &2/5}\bmat{-2/3&1\\1&1}=\bbI.\]
    Plugging this back into the formula, we obtain 
    \begin{align*}
        \E\sparenv{\bfr_{S(n)}}&=\frac{1}{5+n}\bmat{-3/5&3/5\\ 3/5 &2/5}\bmat{\frac{4!}{(n+4)!}\prod_{j=0}^{n-1}(\frac{7}{6}+j+4)&0\\ 0&\frac{4!}{(n+4)!}\prod_{j=0}^{n-1}(2+j+4)}\bmat{-2/3 & 1 \\ 1 & 1}\bmat{2\\ 3} \\ 
        &= \frac{1}{5+n}\bmat{-3/5&3/5\\ 3/5 &2/5}\bmat{\prod_{j=0}^{n-1}\frac{7/6+j+4}{j+5}&0\\ 0&\frac{n+5}{5}}\bmat{5/3\\ 5}\\ 
        &= \frac{1}{5+n}\bmat{-3/5&3/5\\ 3/5 &2/5}\bmat{\frac{5}{3}\prod_{j=1}^{n}\frac{7/6+j+3}{j+4}\\ n+5}. 
    \end{align*} 
    Using the definition of the Gamma function, we obtain 
    \begin{align*}
        \E\sparenv{\bfr_{S(n)}}&=\frac{1}{5+n}\bmat{-3/5&3/5\\ 3/5 &2/5}\bmat{\frac{5}{3}\frac{4!\Gamma(n+3+7/6)}{\Gamma(4+7/6)\Gamma(n+4)}\\ n+5}.
    \end{align*} 
    From Stirling's approximation \eqref{eq:gamma_simple_approx} we have 
    \begin{align*}
        \E\sparenv{\bfr_{S(n)}}&=\frac{1}{5+n}\bmat{-3/5&3/5\\ 3/5 &2/5}\bmat{\frac{5}{3}\frac{4!(n+4)^{1/6}\Gamma(n+4)}{\Gamma(4+7/6)\Gamma(n+4)}\\ n+5}. 
    \end{align*}
    Since $\Gamma(4+7/6)$ is finite, we get, 
     \[\E\sparenv{\bfr_{S(n)}}= \bmat{\frac{3}{2}+o(1)\\ \frac{2}{5}+o(1)}\to \bmat{3/5\\ 2/5} .\] 
\end{example} 

\begin{IEEEproof}[Proof of \Tref{th:basic1}]
Since the mutation law is an $\tau$-mutation law, we know that after each mutation step, the length of the word is increased by $\tau-1$ (we replace a symbol with a word of length $\tau$). Therefore, after $n$ mutation steps the length of the word $\vt^n(w)$ is $|\vt^n(w)|=m+n(\tau-1)$.  
We prove the theorem by induction. 
Let $w\in\cAS$ with $|w|=m$. For the base of induction, we claim that  
\[\E[\bct_{\vt(w)}]=\frac{1}{m}\parenv{\bbM+(m-1)\bbI}\bct_w.\] 
To see this, fix $i\in [d]$ and notice that by definition,  
\[\E\sparenv{\ct_{\vt(w)}(a_i)}=\sum_{\eta\in\cAS} \Pr(\vt(w)=\eta)\ct_{\eta}(a_i).\]
The probability of selecting a symbol $a_{t}$ in the first mutation step is $\frac{\ct_{w}(a_t)}{m}$. 
Moreover, if after the mutation step, $w$ is transformed into $\vt(w)$ by transforming the symbol $a_{t}$ into a word $\eta\in \cAS$, then 
$\ct_{\vt(w)}(a_i)=\ct_{w}(a_i)+\ct_{\eta}(a_i)-\bo_{[t=i]}$. 
Hence we can write $\E\sparenv{\ct_{\vt(w)}(a_i)}$ as 
\[\sum_{t\in [d]}\frac{\ct_{w}(a_t)}{m}\sum_{\eta\in\cAS}\Pr(\vt(a_{t})=\eta)\parenv{\ct_{w}(a_i)+\ct_{\eta}(a_i)-\bo_{[t=i]}}.\]
Opening the parenthesis we get 
\begin{align} 
&\sum_{t\in [d]}\frac{\ct_{w}(a_t)}{m}\sum_{\eta\in\cAS}\Pr(\vt(a_t)=\eta)\ct_{w}(a_i)  \label{eq:pr1eq1}\\
    &\quad +\sum_{t\in [d]}\frac{\ct_{w}(a_t)}{m}\sum_{\eta\in\cAS}\Pr(\vt(a_t)=\eta)\ct_{\eta}(a_i) \label{eq:pr1eq2}\\
    &\quad -\sum_{t\in [d]}\sum_{\eta\in\cAS}\frac{\ct_{w}(a_t)}{m}\Pr(\vt(a_t)=\eta)\bo_{[i=t]}. \label{eq:pr1eq3}
\end{align}
Since $\ct_w(a_i)$ does not depend on $\eta$ or on $t$, \eqref{eq:pr1eq1} can be written as  
\begin{align*}
\ct_{w}(a_i)\sum_{t\in [d]}\frac{\ct_{w}(a_t)}{m}=\ct_w(a_i),
\end{align*}
where the equality follows since there are $m$ symbols in $w$. 
Equation \eqref{eq:pr1eq2} can be written as  
\[\sum_{t\in [d]}\frac{\ct_{w}(a_t)}{m}\sum_{\eta\in\cAS}\Pr(\vt(a_t)=\eta)\ct_{\eta}(a_i)=\sum_{t\in [d]}\frac{\ct_{w}(a_t)}{m}\bbM_{i,t},\] 
and \eqref{eq:pr1eq3} can be written as  
\begin{align*}
    \sum_{t\in [d]}\sum_{\eta\in\cAS}\frac{\ct_{w}(a_t)}{m}\Pr(\vt(a_t)=\eta)\bo_{[t=i]}&= \frac{\ct_{w}(a_i)}{m}\sum_{\eta\in\cAS}\Pr(\vt(a_{i})=\eta) \\ 
    &= \frac{\ct_{w}(a_i)}{m}.
\end{align*}
Putting everything together, we obtain 
\begin{align*} 
\E[\ct_{\vt(w)}(a_i)] &=\ct_{w}(a_i) -\frac{\ct_{w}(a_i)}{m} + \frac{1}{m}\parenv{\bbM \cdot \bct_{w}}_{i}\\
    &= \frac{1}{m}\parenv{\parenv{\bbM+(m-1)\bbI}\cdot\bct_{w}}_{i}.
\end{align*}
Since this is true for every $i\in[d]$, we have 
\[\E[\ct_{\vt(w)}]=\frac{1}{m}(\bbM+(m-1)\bbI)\cdot \bct_{w}.\] 

We assume correctness for $n-1$ and show the induction step. 
\begin{align}
    \E\sparenv{\ct_{\vt^n(w)}(a_i)}&=\sum_{\eta\in\cAS} \Pr(\vt^n(w)=\eta )\ct_{\eta}(a_i) \nonumber \\
    &= \sum_{u\in \cAS} \Pr(\vt^{n-1}(w)=u ) \sum_{t\in [d]} \frac{\ct_{u}(a_t)}{|u|} \sum_{\eta\in\cAS}\Pr(\vt(a_t)=\eta) \parenv{\ct_{u}(a_i)+\ct_{\eta}(a_i)-\bo_{[t=i]}} \nonumber \\
    &= \sum_{u\in \cAS} \Pr(\vt^{n-1}(w)=u ) \sum_{t\in [d]}\frac{\ct_{u}(a_t)}{|u|} \sum_{\eta\in\cAS}\Pr(\vt(a_t)=\eta)\ct_{u}(a_i)  \label{eq:pr1eq11}\\
    &\quad +\sum_{u\in \cAS} \Pr(\vt^{n-1}(w)=u)\sum_{t\in [d]} \frac{\ct_{u}(a_t)}{|u|} \sum_{\eta\in\cAS}\Pr(\vt(a_t)=\eta) \ct_{\eta}(a_i) \label{eq:pr1eq21} \\
    &\quad -\sum_{u\in \cAS} \Pr(\vt^{n-1}(w)=u)\sum_{t\in [d]}\sum_{\eta\in\cAS}\frac{\ct_{u}(a_t)}{|u|} \Pr(\vt(a_t)=\eta)\bo_{[t=i]} \label{eq:pr1eq31}. 
\end{align}
Dealing with each of the arguments explicitly, we may write \eqref{eq:pr1eq11} as 
\begin{align*}
&\sum_{u\in \cAS} \Pr(\vt^{n-1}(w)=u) \ct_{u}(a_i)\sum_{t\in [d]}\frac{\ct_{u}(a_t)}{|u|}\\ 
&=\sum_{u\in \cAS} \Pr(\vt^{n-1}(w)=u) \ct_{u}(a_i) \\ 
&=\E\sparenv{\ct_{\vt^{n-1}(w)}(a_i)},
\end{align*}
Equation \eqref{eq:pr1eq21} can be written as 
\begin{align*}
&\sum_{u\in \cAS} \Pr(\vt^{n-1}(w)=u)\sum_{t\in [d]}\frac{\ct_{u}(a_t)}{|u|}\bbM_{i,t} \\ 
&= \parenv{\bbM\cdot\E\sparenv{\frac{1}{|\vt^{n-1}(w)|}\bct_{\vt^{n-1}(w)}}}_i,
\end{align*}
and \eqref{eq:pr1eq31} can be written as 
\begin{align*}
&\sum_{u\in \cAS} \Pr(\vt^{n-1}(w)=u) \frac{\ct_{u}(a_i)}{|u|}\sum_{\eta\in\cAS}\Pr(\vt(a_{i})=\eta)\\ 
&=\E\sparenv{\frac{1}{|\vt^{n-1}(w)|}\ct_{\vt^{n-1}(w)}(a_i)}.
\end{align*}
Putting everything together with the length $|\vt^{n-1}|=m+(n-1)(\tau-1)$, we have 
\begin{align*}
    \E\sparenv{\ct_{\vt^n(w)}(a_i)}= \frac{1}{m+(n-1)(\tau-1)}\parenv{(\bbM+((n-1)(\tau-1)-1)\bbI)\cdot \E[\ct_{\vt^{n-1}(w)}]}_i.
\end{align*}

Since this is true for all $i\in [d]$, we obtain 
\[\E\sparenv{\bct_{\vt^n(w)}}=\frac{1}{m+(n-1)(\tau-1)}(\bbM+(m+(n-1)(\tau-1)-1)\bbI)\cdot \E[\bct_{\vt^{n-1}(w)}].\] 
From the induction hypothesis, we have  
\[\E\sparenv{\ct_{\vt^{n-1}(w)}}=\prod_{j=0}^{n-2}\frac{1}{m+j(\tau-1)}(\bbM+(m+j(\tau-1)-1)\bbI)\cdot \bct_{w}.\] 
Plugging this into $\E\sparenv{\bct_{\vt^n(w)}}$ yields 
\begin{align*}
    &\E\sparenv{\bct_{S(n)}}\\
    %&= \frac{1}{m+(n-1)(\tau-1)}(\bbM+(m+(n-1)(\tau-1)-1)I)\cdot \E[\bct_{\vt^{n-1}(w)}] \\ 
    %&= \frac{1}{m+(n-1)(\tau-1)}(\bbM+(m+(n-1)(\tau-1)-1)I) \prod_{j=0}^{n-2}\frac{1}{m+j(\tau-1)}(\bbM+(m+j(\tau-1)-1)I)\cdot \bct_w\\ 
    &= \prod_{j=0}^{n-1}\frac{1}{m+j(\tau-1)}(\bbM+(m+j(\tau-1)-1)\bbI)\cdot \bct_w.
\end{align*}
The proof follows since  
\[\E\sparenv{\bfr_{S(n)}}=\frac{1}{m+n(\tau-1)}\E\sparenv{\bct_{\vt^n(w)}}.\]
\end{IEEEproof} 

Our next objective is to generalize Theorem \ref{th:basic1} to incorporate $k$-tuples rather than individual symbols. 
To achieve this, we generalize the definition of the substitution matrix to incorporate $k$-tuples. 
\begin{definition} 
\label{def:k-sub_mat}
    Let $S$ be an $\tau$-mutation system over the alphabet $\cA=\mathset{1_0,\dots,a_{d-1}}$, and let $k\in\N$. 
    The $k$-\textbf{substitution matrix} $\bbM^{(k)}:=\bbM^{(k)}_{(\vt,\bbP)}$ is a real $d^k\times d^k$ matrix, with non-negative entries. 
    The coordinates in $\bbM^{(k)}$ correspond to $k$-tuples, ordered according to the lexicographic order. 
    For $u=(u_0 u_1 \dots u_{k-1}),v=(v_0 v_1 \dots v_{k-1})\in\cA^k$, the $(u,v)$ entry $\bbM^{(k)}_{u,v}$ is  
    \begin{align}
    \label{eq:k_sub_mat}
        \bbM^{(k)}_{u,v}&:= \sum_{\eta\in\cA^\tau}\sum_{t\in [d]}\Pr(\vt(a_t)=\eta)\parenv{\sum_{j=1}^{k-1}\bo_{[v_j=a_t]}\bo_{[(v_0^{j-1}\eta v_{j+1}^{k-1})_{[k]}=u]} +\sum_{j=0}^{\tau-1}\bo_{[v_0=a_t]}\bo_{[\eta_j^{\tau-1} v_1^{k-\tau+j}=u]}}
    \end{align} 
    where $v_0^{j-1}=(v_0\dots v_{j-1})$, $v_{j+1}^{k-1}=(v_{j+1}\dots v_{k-1})$, and $(v_0^{j-1} \eta v_{j+1}^{k-1})_{[k]}$ is the word obtained by taking the first $k$ symbols from $v_0^{j-1} \eta v_{j+1}^{k-1}$.
    When $k$ is clear from the context, we write $\bbM$ instead of $\bbM^{(k)}$.
\end{definition} 

\begin{remark}
    For $k=1$, we have $\bbM^{(1)}=\E[\ct_{\vt(v)}(u)]$ for $u,v\in\cA$. Thus, both Definition \ref{def:k-sub_mat} and Definition \ref{def:sub_mat} coincide.
\end{remark}

\begin{example}
    Consider Example \ref{ex:running} with $\cA=\mathset{0,1}$, and observe that $\vt$ is a $2$-mutation law. 
    We construct the first column in $\bbM^{(2)}$. Initially, we explore the two potential outcomes of $\vt(00)$ when replacing the first coordinate. 
    We derive $000$ with a probability of $2/3$ and $010$ with a probability of $1/3$. 
    According to the formula, we consider only the first two pairs, meaning we count the pairs in a non-cyclic manner. 
    By ordering the coordinates of the vector based on the binary representation, we obtain: 
    \[\frac{2}{3}\bmat{ 2 \\ 0\\ 0\\ 0} + \frac{1}{3}\bmat{0\\1\\1\\0}.\]
    We now consider the two potential outcomes of $\vt(00)$ when replacing the second coordinate. 
    We derive $000$ with a probability of $2/3$ and $001$ with a probability of $1/3$. 
    In this scenario, as per the formula, we only count the first pair that appears. 
    Thus, we obtain: 
    \[\frac{2}{3}\bmat{ 1\\0\\0\\0} + \frac{1}{3}\bmat{1\\0\\0\\0}.\]
    Putting everything together, the first column of $\bbM^{(2)}$ is $\bmat{7/3 \\ 1/3\\1/3\\0}$. 

    Following the same steps for the rest of the columns of $\bbM^{(2)}$, we get 
    \[\bbM^{(2)}=\frac{1}{12}\bmat{28& 11& 6& 3\\ 4& 21& 0& 3\\ 4& 0& 21& 3\\ 0& 4& 9& 27}.\]
\end{example}

Similar to Theorem \ref{th:basic1}, the relation between the $k$-substitution matrix and the frequency of $k$-tuples is demonstrated next.  
\begin{theorem}
    \label{th:basic2}
    Let $S$ be an $\tau$-mutation system over the alphabet $\cA=\mathset{a_0,\dots,a_{d-1}}$ with a starting word $w$ of length $|w|=m$. 
    Fix $k\in \N$ and assume that $m\geq k$. Then the expected frequency of $k$-tuples is given by 
    \begin{align*}
        \E\sparenv{\bfr^{(k)}_{S(n)}}&=\frac{\E\sparenv{\bct^{(k)}_{\vt^n(w)}}}{m+n(\tau-1)} \\ 
        &= \frac{1}{m+n(\tau-1)}\prod_{j=0}^{n-1} \frac{1}{m+j(\tau-1)}\parenv{\bbM^{(k)} +(m+j(\tau-1)-k)\bbI}\bct_w.
    \end{align*}
\end{theorem}

The length of each mutation step is fixed. 
Therefore, Theorem \ref{th:basic2} follows immediately from the following lemma using induction. 
\begin{lemma}
    \label{lem:count_of_next_step_mut} 
    Let $S$ be a mutation system over the alphabet $\cA=\mathset{a_0,\dots,a_{d-1}}$. 
    Fix $k$ and let $w$ be a word of length $|w|\geq k$. Then 
    \[\E\sparenv{\bct^{(k)}_{\vt(w)}}=\frac{1}{|w|}\parenv{\bbM^{(k)}+(|w|-k)\bbI}\cdot \bct_{w}.\] 
\end{lemma} 
The proof of Lemma \ref{lem:count_of_next_step_mut} is technical, thus it is deferred to the appendix. 

\begin{remark}
\label{rem:M_works_in_general}
    Lemma \ref{lem:count_of_next_step_mut} applies to a general mutation system, and is not limited solely to $\tau$-mutation systems. 
    However, the assumption of a constant mutation length is crucial when using induction to prove Theorem \ref{th:basic2}. 
    Without it, the length of $\vt^n(w)$ depends on prior mutation steps, preventing the extraction of the random word's length from the expected value.
\end{remark}

\begin{example}
    \label{ex:running2}
    Let us continue with Example \ref{ex:running}. 
    This time, taking $k=2$ and constructing the substitution matrix $\bbM^{(2)}$ we obtain 
    \[\bbM^{(2)}=\frac{1}{12}\begin{bmatrix} 
        28 & 11   & 6   & 3\\ 
        4  & 21   & 0   & 3\\ 
        4  & 0    & 21  & 3\\ 
        0  & 4    & 9   & 27
    \end{bmatrix}.\] 
    Notice that since $\vt$ is a $2$-mutation and since $k=2$, every mutation on a pair of symbols generates $3$ pairs. 
    Thus, the sum of every column in $\bbM^{(2)}$ adds to $m+k-1=3$. 
    The matrix $\bbM^{(2)}$ has 4 distinct eigenvalues: 
    $\lambda_0=3, \lambda_1=\frac{13}{6} , \lambda_2=\frac{7}{4}, \lambda_3=\frac{7}{6}$. 
    Thus, $\bbM^{(2)}$ can be written as $\bbM^{(2)}=\bbA \Lambda \bbA^{-1}$ where $\Lambda$ is the diagonal matrix containing the eigenvalues, and 
    \[\bbA= \begin{bmatrix}
        \frac{24}{13}   & -\frac{11}{13}   & -\frac{3}{4}   & 1\\ 
        \frac{9}{13}    & -\frac{1}{13}    & \frac{3}{4}    & -1\\ 
        \frac{9}{13}    & -\frac{1}{13}    & -1             & -1\\ 
        1               & 1                & 1              & 1
    \end{bmatrix}.\] 
    Plugging this to $\E\sparenv{\bfr_{S(n)}}$, for large $n$, we obtain 
    \begin{align*}
        \E\sparenv{\bfr_{S(n)}}&= \frac{1}{5+n}\bbA \bbD \bbA^{-1}\bct_w\\
    \end{align*} 
    where $\bbD$ the diagonal matrix obtained by plugging $\Lambda$ into $\prod_{j=0}^{n-1} \frac{1}{m+j(\tau-1)}\parenv{\Lambda +(m+j(\tau-1)-k)I}$. 
    For $i=0,1,2,3$, the $(i,i)$ entry in $\bbD$ is 
    \[\bbD_{i,i}=\frac{4!}{(n+4)!}\prod_{j=0}^{n-1}(\lambda_i+j+3).\] 
    
    Since $\lambda_0=3$ we have 
    \[\frac{4!}{(n+4)!}\prod_{j=0}^{n-1}(\lambda_0+j+3)=\frac{4!(n+5)}{5!}=\frac{(n+5)}{5}.\] 
    For $\lambda_1$, applying Stirling's approximation \eqref{eq:gamma_simple_approx} we get 
    \[\frac{4!}{(n+4)!}\prod_{j=0}^{n-1}(\lambda_1+j+3)=\frac{4!\Gamma(n+4+1/6)}{\Gamma(n+4)\Gamma(5+1/6)}=\frac{4!(n+4)^{1/6}}{\Gamma(31/6)},\] 
    and since $|\lambda_2|,|\lambda_3|<2$ we have that $\frac{1}{(n+1)!}\prod_{j=0}^{n-1}(\lambda_i+j)\to 0$ for $i=2,3$. 
    Considering the starting word $w=00111$, we have $\bct_w=\bmat{1\\1\\1\\2}$. Plugging this into the equation above, for large $n$, we get 
    \[\E\sparenv{\bfr_{\vt^n(w)}}\approx(0.4364, 0.1636, 0.1636, 0.2364).\] 
\end{example}

The example above suggests that in certain instances, the limiting \underline{expected} frequency of $k$-tuples is independent of the 
initial word $w$. 
In contrast, it is clear that the frequency of $k$-tuples (not the expected frequency), when converges, may depend on the starting word, as illustrated by the following example: consider a mutation system $(\vt,\bbP)$ over the binary alphabet, defined as $\vt(0)=00$ and $\vt(1)=11$. 
In this scenario, the convergence of symbol frequencies is almost sure (refer to \cite{elishco2019entropy}), but the limit depends on the choice of the starting word $w$.

Consequently, our objective is to pinpoint situations where the selection of the initial word $w$ has no impact on the expected frequency of $k$-tuples. 
As we show, this condition is satisfied when $\bbM^{(k)}$ is an irreducible matrix or, more accurately, when $\bbM^{(k)}$ has a unique real eigenvalue equal to $k+l-1$. 
This is formally stated in the following claim. 
\begin{claim}
\label{cl:gotoRgen}
    Let $S$ be an $\tau$-mutation system over an alphabet $\cA$ of size $d$ with a starting word $w$ of length $|w|=m$. 
    Let $k$ be a positive integer and let $\bbM=\bbM^{(k)}$ be the $k$-substitution matrix (assume $m\geq k$). 
    Assume $\bbM^{(k)}$ has $s$ \underline{real} eigenvalues $\lambda_0,\dots,\lambda_{s-1}$ that are all equal to $k+\tau-1$, i.e., 
    $\lambda_0=\lambda_1=\dots=\lambda_{s-1}=k+\tau-1$. 
    For $j\in [s]$, let $\bbl_j,\bbr_j$ denote left and right eigenvectors that correspond to $\lambda_j$, normalized such that for $j\in [s]$, $\bbr_j$ is a probability distributions and $\bbl_j\cdot \bbr_j=1$. 
    Denote by $\alpha_j,\; j\in [s]$ the projection, with respect to the standard inner product, of $\bct_w$ on $\bbl_j$, that is $\alpha_j=\langle \bbl_j,\bct_w\rangle$.
    Then, as $n\to\infty$, 
    \[\E\sparenv{\bfr^{(k)}_{S(n)}} \to \frac{1}{m}\sum_{j=0}^{s-1}\alpha_j \bbr_j.\] 
\end{claim} 

We remark that when $s=1$, the left eigenvector associated with $\lambda_0=k+\tau-1$ is the vector of all-ones $\bbl_0=\1$. 
Thus, the projection $\alpha_0$ is given by $\alpha_0=|\bct_w|=m$. 
In this case, the claim suggests that the expected frequency of $k$-tuples is $\bbr_j$, which is independent of $\bct_w$. 

Before proving Claim \ref{cl:gotoRgen} we need a few lemmas. 
The first lemma identifies the spectrum of a $k$-substitution matrix of a $\tau$-mutation system.  
\begin{lemma}
    \label{lem:sum_of_column_l_mut}
    Let $S$ be an $\tau$-mutation system over the alphabet $\cA=\mathset{a_0,\dots,a_{d-1}}$ of size $|\cA|=d$, and let $k\in\N$. 
    Let $\1$ be the (row) vector containing $1$ in every position. 
    Then $\1\cdot \bbM^{(k)}=\tau+k-1$. 
    In other words, the sum of the entries in every column in $\bbM^{(k)}$ is equal to $\tau+k-1$.
\end{lemma} 
The proof of Lemma \ref{lem:sum_of_column_l_mut} is deferred to Section \ref{sec:Avg_l_mut_sys} (see Lemma \ref{lem:sum_of_column}), in which it is proved under more general settings.

The next lemma is used in the proof of Claim \ref{cl:gotoRgen}. 
\begin{lemma}
    \label{lem:normto0}
    Let us denote by $\lambda_0$ the eigenvalue $\rho(\bbM^{(k)})$. 
    \begin{enumerate}
        \item Let $\lambda_i$ be an eigenvalue with $|\lambda_i|< \lambda_0$ and let $\Lambda_i$ the Jordan block that corresponds to $\lambda_i$. 
    Then as $n\to \infty$, 
\[\left\|\frac{1}{m+n(\tau-1)}\prod_{j=0}^{n-1}\frac{\Lambda_i+(m+j(\tau-1)-k)\bbI}{m+j(\tau-1)}\right\|_{\infty}\to \0,\] 
which implies that every entry in $\frac{1}{m+n(\tau-1)}\prod_{j=0}^{n-1}\frac{\Lambda_i+(m+j(\tau-1)-k)\bbI}{m+j(\tau-1)}$ approaches $0$ as $n$ grows.

        \item Let $\lambda_i$ be a real eigenvalue $\lambda_i\leq 0$ and let $\Lambda_i$ the Jordan block that corresponds to $\lambda_i$. Then 
    \[\prod_{j=0}^{n-1}\frac{1}{m+j(\tau-1)}(\Lambda_i+(m+j(\tau-1)-1)\bbI)\to \0\] 
    as $n\to\infty$.
    \end{enumerate}
\end{lemma}

\begin{IEEEproof} 
    Clearly, we have $\rho(\Lambda_i)=\lambda_i$. 
    For every $\epsilon>0$ there exists a matrix norm $\|\cdot \|$ such that for every matrix $\Lambda$, 
    \[\rho(\Lambda) \leq \|\Lambda\|\leq \rho(\Lambda)+\epsilon,\] 
    and $\|\bbI\|=1$ (see, \cite[Lemma 5.6.10]{horn2012matrix}). 
    denote $|\lambda_i|=k+\tau-1-\delta$ with $\delta> 0$, and take $\epsilon=\delta/2$.
    We have 
\begin{align*}
    \prod_{j=0}^{n-1}\frac{\|\Lambda_i+(m+j(\tau-1)-k)I\|}{m+j(\tau-1)}&< \prod_{j=0}^{n-1}\frac{|\lambda_i|+\frac{\epsilon}{2}+m+j(\tau-1)-k}{m+j(\tau-1)}\\ 
    &= \prod_{j=0}^{n-1}\frac{\tau-1-\frac{\delta}{2}+m+j(\tau-1)}{m+j(\tau-1)}.
\end{align*}
This implies that 
\begin{align*}
    \left\|\frac{1}{m+n(\tau-1)}\prod_{j=0}^{n-1}\frac{\Lambda_i+(m+j(\tau-1)-k)\bbI}{m+j(\tau-1)}\right\| &< \frac{1}{m+n(\tau-1)}\prod_{j=0}^{n-1}\frac{\tau-1-\frac{\delta}{2}+m+j(\tau-1)}{m+j(\tau-1)}\\ 
    &=\frac{1}{m}\prod_{j=1}^{n}\frac{m+j(\tau-1)-\delta/2}{m+j(\tau-1)}
\end{align*}
which vanishes as $n\to\infty$. 
In a finite-dimensional space, all norms are equivalent, thus 
\[\left\|\prod_{j=0}^{n-1}\frac{1}{m+j(\tau-1)}(\Lambda_i+(m+j(\tau-1)-1)\bbI)\right\|_{\infty}\to 0.\] 

The proof of the second part is similar. 
Let us denote $\gamma=\floorenv{\frac{|\lambda_i|}{\tau-1}}$, and for $n$ large enough, write 
    \begin{align*}
        &\prod_{j=0}^{n-1}\frac{1}{m+j(\tau-1)}(\Lambda_i+(m+j(\tau-1)-1)\bbI) \\ 
        &= \parenv{\prod_{j=0}^{\gamma-1}\frac{\Lambda_i+(m+j(\tau-1)-1)\bbI}{m+j(\tau-1)}} 
        \prod_{j=\gamma}^{n-1}\frac{\Lambda_i+(m+j(\tau-1)-1)\bbI}{m+j(\tau-1)}.
    \end{align*}
    We have that 
    \[\prod_{j=\gamma}^{n-1}\frac{\Lambda_i+(m+j(\tau-1)-1)\bbI}{m+j(\tau-1)}=\prod_{j=0}^{n-\gamma-1}\frac{\Lambda_i+(m+\gamma (\tau-1)+j(\tau-1)-1)\bbI}{m+\gamma (\tau-1)+j(\tau-1)}.\]
    Notice that $-1\leq \frac{\lambda_i}{\tau-1}+\gamma\leq 0$ and that  
    \[\prod_{j=0}^{n-\gamma-1}\frac{\Lambda_i+(m+\gamma (\tau-1)+j(\tau-1)-1)\bbI}{m+j(\tau-1)}= \prod_{j=0}^{n-\gamma-1}\frac{\parenv{\frac{\Lambda_i}{\tau-1}+\gamma \bbI}+(\frac{m-1}{\tau-1}+j)\bbI}{\frac{m}{\tau-1}+j}.\]  
    Thus, the values on the diagonal of $\parenv{\frac{\Lambda_i}{\tau-1}+\gamma \bbI}+(\frac{m-1}{\tau-1}+j)\bbI$ are smaller than $\frac{m}{\tau-1}+j$. 
    This, together with the fact that $\gamma$ is finite, imply the result. 
\end{IEEEproof}

We are now ready to prove Claim \ref{cl:gotoRgen}. 
\begin{IEEEproof}[Proof of Claim \ref{cl:gotoRgen}]
    Let us write $\bbM^{(k)}=\bbA\Lambda \bbA^{-1}$ where $\Lambda$ is in Jordan canonical form. 
    For simplicity, we denote the eigenvalues of $\bbM^{(k)}$ as $\lambda_0,\lambda_1,\dots, \lambda_{r-1}$ with $r\leq d^k$, ordered in a decreasing order of modulus 
    $\lambda_0=k+\tau-1\geq |\lambda_1|\geq |\lambda_2|\geq \cdots \geq |\lambda_{r-1}|$. 
    From Theorem \ref{th:non_neg_mat_properties} we deduce that if $|\lambda_j|=k+\tau-1$ then there exists a unique eigenvector $\bbr_j$ (and not a generalized eigenvector) such that $\bbM^{(k)}\cdot \bbr_j=\lambda_j \bbr_j$. 

    Let us consider now an eigenvalue $\lambda=\lambda_i$ with $|\lambda|=k+\tau-1$. 
    Since $\lambda$ is of maximum modulus, Theorem \ref{th:non_neg_mat_properties} suggests that its corresponding Jordan block is one-by-one. 
    Assume $\Lambda_i$ is the Jordan block associated with $\lambda$, and that $\Lambda$ contains $\lambda_i$ in the $(i,i)$ coordinate. 
    Plugging $\bbA\Lambda \bbA^{-1}$ into the formula of $\bfr_{S(n)}$, we get 
    \[\E\sparenv{\bfr_{S(n)}}=\frac{1}{m+n(\tau-1)}\bbA \bbD \bbA^{-1}\bct_w\] 
    where $\bbD_{i,i}$ contains $\prod_{j=0}^{n-1} \frac{\lambda+m-k+j(\tau-1)}{m+j(\tau-1)}$. 
    Calculating the product, we get 
    \begin{align*}
        &\prod_{j=0}^{n-1} \frac{\lambda+m-k+j(\tau-1)}{m+j(\tau-1)}\\ 
        &= \prod_{j=0}^{n-1} \frac{\frac{\lambda+m-k}{\tau-1}+j}{\frac{m}{\tau-1}+j}\\ 
        &= \frac{\Gamma\parenv{\frac{\lambda+m-k}{\tau-1}+n}}{\Gamma\parenv{\frac{\lambda+m-k}{\tau-1}}} 
        \frac{\Gamma\parenv{\frac{m}{\tau-1}}}{\Gamma\parenv{\frac{m}{\tau-1}+n}}\\ 
        &\stackrel{(a)}{=} n^{\frac{\lambda-k}{\tau-1}}\frac{\Gamma\parenv{\frac{m}{\tau-1}}}{\Gamma\parenv{\frac{\lambda+m-k}{\tau-1}}}
    \end{align*}
    where $(a)$ follows from Stirling's approximation \eqref{eq:gamma_simple_approx}. 
    If $\lambda$ is complex, we may write $\lambda=\lambda^r+\bbi\lambda^c$ where $\lambda^r,\lambda^c\in\R$ are its real and imaginary parts, respectively. 
    We obtain 
    \[\prod_{j=0}^{n-1} \frac{\lambda+m-k+j(\tau-1)}{m+j(\tau-1)}=n^{\frac{\lambda^r-k}{\tau-1}}n^{\bbi\frac{\lambda^c}{\tau-1}}\frac{\Gamma\parenv{\frac{m}{\tau-1}}}{\Gamma\parenv{\frac{\lambda+m-k}{\tau-1}}}.\]
    For any real number $\alpha\in \R$, 
    \begin{align*}
        n^{\bbi\alpha}&= e^{\log \parenv{n^{\bbi\alpha}}}\\ 
        &= e^{\bbi\alpha\log (n)}\\ 
        &= \cos \parenv{\alpha\log(n)}+\bbi\sin\parenv{\alpha\log(n)}.
    \end{align*} 
    Thus, $\abs{n^{\bbi\frac{\lambda^c}{\tau-1}}}$ is bounded, and $\frac{n^{\bbi\frac{\lambda^c}{\tau-1}}}{m+n(\tau-1)}\to 0$ as $n$ grows\footnote{In fact, it is possible to show that the imaginary part cancels out when calculating $\E[\bfr_{S(n)}]$. This follows since $\bbM^{(k)}$ is real. Thus, if $\bbv$ is an eigenvector associated with $\lambda$, then $\overline{\bbv}$ is an eigenvector associated with $\overline{\lambda}$, and $|\overline{\lambda}|=|\lambda|$.}. 
    In addition, if $\lambda=\lambda^r+\bbi\lambda^c$ with $|\lambda^c|>0$, then the maximum modulus $|\lambda|=k+\tau-1$ implies that $\lambda^r<k+\tau-1$. 
    This means that $n^{\frac{\lambda^r-k}{\tau-1}}=n^{1-\epsilon}$ for some $\epsilon>0$, which in turn, implies that 
    \[\frac{n^{\frac{\lambda^r-k}{\tau-1}}}{m+n(\tau-1)}\to 0.\]
    Using Lemma \ref{lem:normto0} and the discussion above, we obtain that $\E[\bfr_{S(n)}]$ is affected only by the real eigenvalues that are equal to $k+\tau-1$ and their corresponding eigenvectors. 
    In other words, when estimating $\E[\bfr_{S(n)}]$ for large $n$, we may ignore all eigenvalues and eigenvectors except for the \underline{real}, maximum modulus eigenvalues and their eigenvectors.
    
    Let us denote by $s$ the number of real eigenvalues that are equal to $k+\tau-1$. 
    Let $\bbA_{[s]}$ be the $k\times s$ matrix obtained by taking the $s$ columns of $\bbA$ that correspond to the maximum modulus eigenvectors, 
    and let $\bbA^{-1}_{[s]}$ be the $s\times k$ matrix obtained by taking the $s$ rows of $\bbA^{-1}$ that correspond to the maximum modulus eigenvectors. 
    Notice that $\bbA_{[s]} \bbA^{-1}_{[s]}=\bbI^{s\times s}$. 
    Consider also the $s\times s$ diagonal matrix $\Lambda_{[s]}$ that contains the real maximal eigenvalues from $\Lambda$ (so $\Lambda_{[s]}$ contains $\tau+k-1$ on its diagonal). 
    Plugging this to the formula of $\bfr_{S(n)}$, we get 
    \[\E\sparenv{\bct_{S(n)}}=\bbA_{[s]} \Lambda_{[s]} \bbA^{-1}_{[s]}\bct_w.\] 
    Noticing that the rows of $\bbA^{-1}$ are the left eigenvectors of $\bbM^{(k)}$, we obtain the claim. 
    According to Corollary \ref{cor:sum_left_eigenvectors_is_1}, $\sum_{j=0}^{s-1}\alpha_j \bbr_j=m\1$, and indeed, 
    the fraction $\frac{1}{m}$ that appears in the formula ensures that the sum is a probability vector. 
\end{IEEEproof}

\begin{example} 
\label{ex:two_eigenvalues}
    Let $(\vt,\bbP)$ be a mutation law over the binary alphabet, defined as $\vt(0)=0^a,\; \vt(1)=1^a$ for some $a\in\N$. 
    Let $w=01$ be the starting word so for $k=2$, we have $\bct_w=\bmat{0\\1\\1\\0}$. 
    Calculating $\bbM^{(2)}$, we obtain 
    \[\bbM^{(2)}=\bmat{a+1&a-1&0&0\\0&2&0&0\\0&0&2&0\\0&0&a-1&a+1}.\] 
    For $a=1$ we have $\bbM^{(2)}=2\bbI$ with the standard basis as eigenvectors. 
    Applying Claim \ref{cl:gotoRgen} we get $\E\sparenv{\bfr^{(2)}_{S(n)}}=\bmat{0\\1/2\\1/2\\0}$ which is quite clear, since when $a=1$ the word $w$ stays the same after every mutation step. 
    For $a>1$, the eigenvalues are $2,2,a+1,a+1$ so two eigenvalues are equal to $k+\tau-1=a+1$. 
    Their corresponding eigenvectors are $\bbl_0=\bmat{1&1&0&0},\bbl_1=\bmat{0&0&1&1}$, so the projections are $\alpha_0=\alpha_1=1$. 
    From Claim \ref{cl:gotoRgen}, we obtain that for large $n$, 
    \[\E\sparenv{\bfr^{(2)}_{S(n)}}=\frac{1}{2}\parenv{\alpha_0\bmat{1\\0\\0\\0}+\alpha_1\bmat{0\\0\\0\\1}}=\bmat{1/2\\0\\0\\1/2}.\]
\end{example} 

Limiting the scope of our study to the subset of $\tau$-mutation systems facilitates the computation of the expected frequency of $k$-tuples at each mutation step. Our next objective is to lift the restriction of a fixed mutation length. 
Relaxing the restriction on fixed mutation length will lead to slightly less robust results.

\section{Average $\tau$-mutation system}
\label{sec:Avg_l_mut_sys}
%%%%%%%%%%%%%%%%%%%%%%%%%%%%%%%%%%%%%
In this section, we explore mutation laws that are slightly more general. 
Instead of adhering strictly to $\tau$-mutation laws, where the length of a word increases by a fixed constant during a mutation step, 
we now only require that, \underline{on average}, each mutation step results in a fixed constant increase in the word's length. 
First, we define an average mutation law.
 
\begin{definition}
    \label{def:ave_tau_mut} 
    Let $(\vt,\bbP)$ be a mutation law over an alphabet $\cA={a_0,\dots,a_{d-1}}$. 
    We say that $(\vt,\bbP)$ is an \textbf{average $\tau$-mutation law} if for every $t\in [d]$, $\E\sparenv{\abs{\vt(a_t)}}=\tau$. 
    We call $S$ an \textbf{average $\tau$-mutation system}, if the mutation law in $S$ is an average $\tau$-mutation law. 
\end{definition}
It is clear that a $\tau$-substitution system is an average $\tau$-substitution system.

Our next goal is to adjust the definition of $k$-substitution matrix to average $\tau$-mutation systems. 
The definition is similar to Definition \ref{def:k-sub_mat}, only that in this case we incorporate the different lengths that a mutation step can add to a word. 
\begin{definition} 
\label{def:k-sub_mat2}
    Let $S$ be an average $\tau$-mutation system over the alphabet $\cA=\mathset{1_0,\dots,a_{d-1}}$ of size $|\cA|=d$, and let $k\in\N$. 
    The $k$-\textbf{substitution matrix} $\bbM^{(k)}:=\bbM^{(k)}_{\vt}$ is a real $d^k\times d^k$ matrix, with non-negative entries, defined as follows. 
    For $u=(u_0 u_1 \dots u_{k-1}),v=(v_0 v_1 \dots v_{k-1})\in\cA^k$, the $(u,v)$ entry in $\bbM^{(k)}$ is  
    \begin{align}
    \label{eq:k_sub_mat2}
        \bbM^{(k)}_{u,v}&:= \sum_{l\in \N}\sum_{\eta\in\cA^l}\sum_{t\in [d]}\Pr(\vt(a_t)=\eta)\parenv{\sum_{j=1}^{k-1}\bo_{[v_j=a_t]}\bo_{[(v_0^{j-1}\eta v_{j+1}^{k-1})_{[k]}=u]} +\sum_{j=0}^{l-1}\bo_{[v_0=a_t]}\bo_{[\eta_j^{l-1} v_1^{k-l+j}=u]}}
    \end{align} 
    where $(v_0^{j-1} \eta v_{j+1}^{k-1})_{[k]}$ is the sequence obtained by taking the first $k$ symbols from $v_0^{j-1} \eta v_{j+1}^{k-1}$.
    When $k$ is clear from the context, we write $\bbM$ instead of $\bbM^{(k)}$.
\end{definition} 

\begin{example} 
\label{ex:mos} 
        The following example is taken from \cite{lou2019evolution} (see also Figure 2 in their paper). 
        Let $(\vt,\bbP)$ be a mutation law over the binary alphabet defined as follows. 
        \[\vt(0)=\begin{cases} 1& \text{w.p. } \alpha\\ 00&\text{w.p. } (1-\alpha)\end{cases}, \;  \vt(1)=\begin{cases} 0& \text{w.p. } \alpha\\ 11&\text{w.p. } (1-\alpha)\end{cases}.\] 
        It is straightforward to check that this is an average $(7/4)$-mutation system. We calculate $\bbM^{(2)},\bbM^{(3)}$ and obtain 
        \[\bbM^{(2)}= \bmat{3(1-\alpha)&1&\alpha&0\\ \alpha&2(1-\alpha)&0&\alpha\\ \alpha&0&2(1-\alpha)&\alpha\\0&\alpha&1&3(1-\alpha)},\] 
        and 
        \[\bbM^{(3)}=\bmat{4(1-\alpha)&2-\alpha&\alpha&0&\alpha&0&0&0\\\alpha&2(1-\alpha)&1-\alpha&1&0&\alpha&0&0\\\alpha&0&2(1-\alpha)&\alpha&0&0&\alpha&0\\0&\alpha&1&3(1-\alpha)&0&0&0&\alpha\\ \alpha&0&0&0&3(1-\alpha)&1&\alpha&0\\0&\alpha&0&0&\alpha&2(1-\alpha)&0&\alpha\\ 0&0&\alpha&0&1&1-\alpha&2(1-\alpha)&\alpha\\ 0&0&0&\alpha&0&\alpha&2-\alpha&4(1-\alpha)}.\]
\end{example}

From Definition \ref{def:k-sub_mat2}, we immediately obtain the following property. 
\begin{lemma}
    \label{lem:sum_of_column}
    Consider an average $\tau$-mutation law $(\vt,\bbP)$ over the alphabet $\cA=\mathset{a_0,\dots,a_{d-1}}$ with a size of $|\cA|=d$, and let $k\in\N$. 
    Let $\1$ be the (row) vector containing $1$ in every position. 
    Then, $\1\cdot \bbM^{(k)}=\tau+k-1$. 
    In simpler terms, the sum of the entries in every column in $\bbM^{(k)}$ is equal to $\tau+k-1$.
\end{lemma} 

\begin{IEEEproof}
    The proof follows through a straightforward calculation. 
    Consider $v\in\cA^k$ and sum $\bbM_{u,v}$ over all $u\in\cA^k$. 
    Note that for a fixed $l,t$,  
    \begin{align*}
        &\sum_{u\in\cA^k}\parenv{\sum_{j=1}^{k-1}\bo_{[v_j=a_t]}\bo_{[(v_0^{j-1}\eta v_{j+1}^{k-1})_{[k]}=u]} +\sum_{j=0}^{l-1}\bo_{[v_0=a_t]}\bo_{[\eta_j^{l-1} v_1^{k-l+j}=u]}}\\ 
        &= \parenv{\ct_v(a_t)+(l-1)\bo_{[v_0=a_t]}}.
    \end{align*} 
   Plugging this into $\sum_{u\in\cA^k}\bbM_{u,v}$, we obtain 
    \[\sum_{u\in\cA^k}\bbM_{u,v}= \sum_{l\in \N}\sum_{\eta\in\cA^l}\sum_{t\in [d]}\Pr(\vt(a_t)=\eta)\parenv{\ct_v(a_t)+(l-1)\bo_{[v_0=a_t]}}.\] 
    Breaking this sum into two parts and addressing the first, we get    
    \begin{align*}
    \sum_{l\in \N}\sum_{\eta\in\cA^l}\sum_{t\in [d]}\Pr(\vt(a_t)=\eta)\ct_v(a_t)&=\sum_{t\in[d]}\ct_v(a_t)\\ 
    &=k.
    \end{align*}
    The second sum is equal to 
    \begin{align*}
    &\sum_{l\in \N}\sum_{\eta\in\cA^l}\sum_{t\in [d]}\Pr(\vt(a_t)=\eta)(l-1)\bo_{[v_0=a_t]}\\
    &=\sum_{t\in [d]}\bo_{[v_0=a_t]}\sum_{l\in \N}(l-1)\sum_{\eta\in\cA^l}\Pr(\vt(a_t)=\eta)\\
    &=\sum_{l\in \N}(l-1)\sum_{\eta\in\cA^l}\Pr(\vt(v_0)=\eta)\\
    &= \sum_{l\in \N}(l-1) \Pr(|\vt(v_0)|=l)\\ 
    &=\sum_{l\in \N}l\Pr(|\vt(v_0)|=l)- 1 \\ 
    &= \tau-1,
    \end{align*} 
    where the last equality follows since $\vt(\cdot)$ is an average $\tau$-mutation law.
\end{IEEEproof}

\begin{remark}
    From Lemma \ref{lem:sum_of_column} and Theorem \ref{th:non_neg_mat_properties}, we deduce that the spectral radius $\rho(\bbM^{(k)})$ of the 
    $k$-substitution matrix of an average $\tau$-mutation law is equal to $\tau+k-1$. Additionally, every Jordan block corresponding to a maximum modulus eigenvalue is of size one. 
    Furthermore, the vector of all ones, $\bbl=\1$ is a left eigenvector associated with the eigenvalue $\lambda_0=\tau+k-1$.
\end{remark}

Recall that, in the analysis of $\tau$-mutation systems, the deterministic length of the random word $S(n)$ obtained after $n$ mutation steps plays a crucial role. In average $\tau$-mutation systems, the length of $S(n)$ becomes random. Accurately estimating the length $S(n)$ will be useful in our analysis. 
Therefore, if $S$ is an average $\tau$-mutation system with a starting word $w$ of length $m$, our first objective is to demonstrate that the length of $S(n)=\vt^n(w)$ is almost surely $n(\tau-1)+m$. To achieve this, we rely on the following established theorem, available in \cite{hall2014martingale}.

\begin{theorem}[{{\cite[Theorem 2.19]{hall2014martingale}}}]
\label{th:Hall}
    Let $\mathset{X_n}$ be a sequence of random variables and let $\mathset{\cF_n}$ be an increasing sequence of $\sigma$-algebras with 
    $X_n$ measurable w.r.t. $\cF_n$ for every $n$. 
    Let $X$ be a random variable and $c$ a constant such that $\E[|X|]<\infty$ and $\Pr(|X_n|>a)\leq c\Pr(|X|>a)$ for every $x\geq 0$ and $n\geq 1$. 
    Assume also that $\E[|X|\max (\log(|X|),0) ]<\infty$. 
    Then, 
    \[\frac{1}{n}\sum_{i=1}^n\parenv{X_i-\E[X_i ~|~ \cF_{i-1}]}\stackrel{a.s}{\rightarrow} 0\] 
    as $n\to\infty$.
\end{theorem}

We now show that the length of $\vt^n(w)$ almost surely converges to $m+n(r-1)$. 
The proof idea is akin to many proofs of this kind that emerge when studying urn models \cite{mahmoud2008polya}.
\begin{lemma} 
\label{lem:limlength} 
Let $S$ be an average $\tau$-mutation system with a starting word $w$ of length $|w|=m$. 
Then almost surely, 
    \[\frac{|S(n)|-m}{n}\to \tau_0-1.\]
\end{lemma}

\begin{IEEEproof}
    Let $Y_n$ denote the (random) length of $S(n)=\vt^n(w)$. 
    Consider $\cF_n$ as the $\sigma$-algebra generated by $\bct_j$ for $j\leq n$. 
    From $\cF_n$, we have knowledge of $\bct_{\vt^n(w)}$, allowing us to derive $\fr_{\vt^n(w)}(a_t)$ for each symbol $a_t$. 
    We have 
    \begin{align*}
        \E[Y_n-Y_{n-1} ~|~ \cF_{n-1}]&= \sum_{t\in [d]}\E\sparenv{\abs{\vt(a_t)} ~|~ \cF_{n-1}}\fr_{\vt^{n-1}(w)}(a_t) -1 \\ 
        &= \tau\sum_{t\in [d]}\fr_{n-1}(a_t) -1\\ 
        &= \tau-1.
    \end{align*} 
    Hence, we conclude that 
    \[\E[Y_n-Y_{n-1} ~|~ \cF_{n-1}]=\tau-1.\] 
    Since the number of mutations possible for a symbol is finite, we introduce the random variable $X$ as the constant value $\max_t \mathset{|\vt(a_t)|}$. 
    It follows that $\E[|X|]<\infty$, and $\Pr(|Y_n-Y_{n-1}|>a)\leq \Pr(|X|>a)$. 
    Applying Theorem \ref{th:Hall}, we obtain 
    \[\frac{1}{n}\sum_{i=1}^n \parenv{Y_i-Y_{i-1}-\E[Y_i-Y_{i-1} ~|~ \cF_{i-1}]}\stackrel{a.s.}{\to} 0.\] 
    Thus, 
    \[\frac{Y_n-Y_0}{n}-(\tau-1) \stackrel{a.s.}{\to}0\] 
    which implies that almost surely 
    \[\frac{Y_n-m}{n}\to \tau-1.\] 
\end{IEEEproof}

Removing the constraint of fixed-length mutations and replacing it with an average length does not permit the examination of the expected frequency of $k$-tuples for every $n$. Instead, it allows the study of the expected frequency in the limit as $n$ approaches infinity, which is the focus of the next subsection. 

\subsection{Expected frequency of $k$-tuples as $n\to\infty$}
%%%%%%%%%%%%%%%%%%%%%%%%%%%%%%%%%%%%%%%%%%%%%%%%%%%%%%%%%%%%%%
In this subsection, we investigate the expected frequency of $k$-tuples in $S(n)$, where $S$ represents an average $\tau$-mutation system and $n$ is sufficiently large. To achieve this, we employ stochastic approximation (for a comprehensive study, see \cite{borkar2009stochastic}). 
We mention here that the use of stochastic approximation is not necessary in our case, and other (non-probabilistic) iterative methods may be used. 
Nevertheless, to remain consistent with previous papers, we use the method of stochastic approximation. 
Before delving into the analysis, we review some fundamental definitions and theorems. 

We say that a set $A$ in $\R^d$ is \textbf{connected} if it is not a union of disjoint sets. 
A set $A$ is called an \textbf{internally chain transitive invariant set} for a differential equation $\dot{\bbx}_t=f(\bbx_t)$ if it is closed, 
if $\bbx_{t_0}\in A$ for some $t_0$ implies that $\bbx_t\in A$ for $t\geq t_0$, and if for every $a,a'\in A$, and positive reals $T,\epsilon$, 
there exists a sequence $a=a_0,a_1,\dots, a_{n-1}=a'$ such that if $\bbx_0=a_i$ for $i<n-1$, then $\exists t\geq T$ such that $\bbx_t$ is $\epsilon$ close to $a_{i+1}$.

The main result of this subsection relies on the following stochastic approximation theorem. 
\begin{theorem}\cite[Section 2]{borkar2009stochastic} 
\label{th:stochastic_approx}
    Let $\mathset{\bbx_n}_n$ be a sequence over $\R^d$ such that for $n\geq 0$, 
    \[\bbx_{n+1}=\bbx_n+a(n)\parenv{h(\bbx_n)+M_{n+1}+\epsilon_n},\]
    with a given $\bbx_0$. 
    Assume also that 
    \begin{enumerate}
    \item The map $h:\R^d\to \R^d$ is Lipschitz. 
    \item $\mathset{M_n}_n$ is a square-integrable martingale. 
    \item $a(n)$ is positive for every $n$, $\sum_{n\geq 0} a(n)=\infty$, and $\sum_{n\geq 0} a(n)^2<\infty$. 
    \item The supremum $\|\bbx_n\|$ is bounded, i.e., $\sup_n \|\bbx_n\|<\infty$ almost surely. 
    \item The sequence $\mathset{\epsilon_n}$ (which may be random) is bounded and $o(1)$.
    \end{enumerate}
    Then almost surely, $\bbx_n$ converges to a compact connected internally chain transitive invariant set of the differential equation 
    \[\dot{\bbx}_t=h(\bbx_t).\]
\end{theorem}

We now state the main result of this subsection, which bears resemblance to \cite[Th. 10]{farnoud2019estimation} which holds for substitution and duplication systems. 
\begin{theorem}
    \label{th:general_expected_freq}
    Consider an average $\tau$-mutation system $S$ over a finite alphabet $\cA=\mathset{a_0,\dots,a_{d-1}}$. 
    Fix $k$, and let $w$ be an initial word with length $|w|=m\geq k$. Suppose $\bbM^{(k)}$ has $s_0$ real eigenvalues $\lambda_0,\dots,\lambda_{s_0-1}$ equal to $\rho(\bbM^{(k)})=\tau+k-1$. For each $j\in [s_0]$, denote $\bbl_j$ and $\bbr_j$ the left and right eigenvectors associated with $\lambda_j=r+k-1$. 
    Normalize $\bbr_j$ to form a probability vector, and normalize $\bbl_j$ such that $\bbl_j\cdot \bbr_j=1$. 
    Define $\alpha_j$ as the projection of $\bct_w$ onto $\bbl_j$, denoted by $\alpha_j=\langle \bbl_j,\bct_w\rangle$. 
    Then, almost surely, the limit $\lim_{n\to\infty} \E\sparenv{\bfr_{S(n)}}$ exists, and 
    \[\lim_{n\to\infty} \E\sparenv{\bfr_{S(n)}}=\frac{1}{m}\sum_{j\in [s_0]}\alpha_j \bbr_j.\]
\end{theorem} 
We remark that Corollary \ref{cor:sum_left_eigenvectors_is_1} implies that the sum $\frac{1}{m}\sum_{j\in [s_0]}\alpha_j \bbr_j$ is indeed a probability vector. 

\begin{example} 
    Let us continue with Example \ref{ex:mos}. 
    The eigenvalues of $\bbM^{(2)}$ are $\mathset{3-\alpha, 3(1-\alpha), 2(1-\alpha), 2(1-2\alpha)}$, and the normalized (right) eigenvector $\bbr$ associated with 
    $\lambda_0=3-\alpha$ is 
    \[\bbr=\frac{1}{2(3\alpha+1)}\bmat{1+\alpha\\ 2\alpha\\ 2\alpha\\ 1+\alpha}.\]
    Thus, Theorem \ref{th:general_expected_freq} states that $\lim_{n\to\infty} \E\sparenv{\bfr^{(2)}_{S(n)}}=\bbr$. 
    In \cite{lou2019evolution} the authors obtained that in this example, $\bfr^{(k)}_{S(n)}$ converges almost surely to the same distribution. 
    The expected frequency of pairs should be the same, as we obtained here. 

    Similarly, the eigenvalues of $\bbM^{(3)}$ are $\mathset{4-\alpha, 2-\alpha,3-2\alpha, 4-3\alpha,2-3\alpha, 2-3\alpha,3-4\alpha,2-5\alpha}$ 
    and the normalized (right) eigenvector $\bbr$ that corresponds to $\lambda_0=4-\alpha$ is 
    \[\bbr^T=\frac{1}{4(1+4\alpha+4\alpha^2)}\bmat{\alpha^2+\alpha+2&\alpha^2+3\alpha&\alpha+3\alpha^2&3\alpha+\alpha^2&3\alpha+\alpha^2&\alpha+ 3\alpha^2&3\alpha+\alpha^2&\alpha^2+\alpha+2}.\]
    Thus, Theorem \ref{th:general_expected_freq} states that $\lim_{n\to\infty} \E\sparenv{\bfr^{(3)}_{S(n)}}=\bbr$.
    Using the result from \cite{lou2019evolution}, the expected frequency of triples is almost surely, the limiting frequency. 
\end{example}

We now prove Theorem \ref{th:general_expected_freq}. The proof is similar to the proof of \cite[Th. 10]{farnoud2019estimation}.
\begin{IEEEproof}[Proof of Theorem \ref{th:general_expected_freq}] 
    For simplicity, we write $\bct_n$ instead of $\bct^{(k)}_{S(n)}$, and write $\bbM$ instead of $\bbM^{(k)}$. 
    Let us denote by $\bbz_n$ the expected vector $\bbz_n=\E\sparenv{\bct_n}$, and let $Y_n$ denote the length of $\vt^n(w)$, $Y_n=|\vt^n(w)|$. 
    According to Lemma \ref{lem:count_of_next_step_mut} and Remark \ref{rem:M_works_in_general}, we can write  
    \begin{align*}
        \bbz_{n+1}&= \E\sparenv{\parenv{\bbM+(Y_n-k)\bbI}\frac{1}{Y_n}\bct_n}\\ 
        &= \bbz_n+ \parenv{\bbM-k\bbI}\E\sparenv{\frac{1}{Y_n}\bct_n}.
    \end{align*}
    We divide by $n(\tau-1)+m$ and use the equality 
    \[\frac{1}{(n+1)(\tau-1)+m}=\frac{1}{n(\tau-1)+m}-\frac{\tau-1}{((n+1)(\tau-1)+m)(n(\tau-1)+m)}\] 
    to obtain 
    \begin{align*}
    &\frac{1}{(n+1)(\tau-1)+m}\bbz_{n+1}=\frac{1}{n(\tau-1)+m}\bbz_n+\\ 
    &\quad \frac{1}{(n+1)(\tau-1)+m} \E\sparenv{(\bbM+k\bbI)\frac{1}{Y_n}\bct_n-\frac{\tau-1}{n(\tau-1)+m}\bct_n}. 
    \end{align*} 
    We now add and subtract $\frac{\bct_n}{n(\tau-1)+m}$ to obtain 
    \begin{align*}
    &\frac{1}{n(\tau-1)+m}\bbz_n+\frac{1}{(n+1)(\tau-1)+m}(\bbM-(\tau+k-1)\bbI)\E\sparenv{\frac{1}{n(\tau-1)+m}\bct_n}\\ 
    &\quad + \frac{\tau-1}{n(\tau-1)+m}(\bbM-(\tau+k-1)\bbI)\E\sparenv{\parenv{\frac{1}{Y_n}-\frac{1}{n(\tau-1)+m}}\bct_n}. 
    \end{align*} 
    We now denote $x_n=\frac{1}{n(\tau-1)+m}\bbz_n$ and obtain 
    \begin{align*}
    \bbx_{n+1}&=\bbx_n+\frac{1}{(n+1)(\tau-1)+m}(\bbM-(\tau+k-1)\bbI)\bbx_n \\ 
    & + \frac{\tau-1}{n(\tau-1)+m}(\bbM-(\tau+k-1)\bbI)\E\sparenv{\parenv{\frac{1}{Y_n}-\frac{1}{n(\tau-1)+m}}\bct_n}. 
    \end{align*}

    We notice that  
    \[\frac{\tau-1}{n(\tau-1)+m}(\bbM-(\tau+k-1)\bbI)\E\sparenv{\parenv{\frac{1}{Y_n}-\frac{1}{n(\tau-1)+m}}\bct_n}=o(1)\] 
    is bounded, and Lemma \ref{lem:limlength} implies that it is indeed $o(1)$. 
    The matrix $\bbM+(\tau+k-1)\bbI$, viewed as a linear transformation, is Lipschitz. 
    Since $\1\cdot \bct_n\leq C(n(\tau-1)+m)$ for some constant $C$, we obtain that $\bbx_n$ is bounded almost surely. 
    Thus, all the conditions in Theorem \ref{th:stochastic_approx} are satisfied, and $\bbx_n$ almost surely follows $\dot{\bbx}_t=(\bbM-(\tau+k-1)\bbI)\bbx_t$. 

    Solving the differential equation we obtain 
    \[\bbx_t=e^{(\bbM-(\tau+k-1)\bbI)t}x_0,\] 
    with $\bbx_0=\frac{1}{m}\bct_w$ and where matrix exponent $e^{\bbA}$ is defined as 
    \[e^{\bbA}=\sum_{j=0}^{\infty} \frac{1}{j!}\bbA^j.\]
    The maximal (real) eigenvalue of $\bbM-(\tau+k-1)\bbI$ is $0$. 
    We now use Jordan decomposition to write 
    \[\bbM-(\tau+k-1)\bbI=\bbP e^\Lambda \bbP^{-1}\] 
    where the rows of $\bbP^{-1}$ are the left eigenvectors (and generalized eigenvectors) of $\bbM-(\tau+k-1)\bbI$. 
    Moreover, from Theorem \ref{th:non_neg_mat_properties}, the Jordan blocks that correspond to maximum modulus eigenvalues are one-by-one. 

    Assume $\Lambda_i$ is a Jordan block in $\Lambda$, associated with the eigenvalue $\lambda_i$. 
    Then $\Lambda_i=\text{Diag}(\lambda_i)+N$ where $\text{Diag}(\lambda_i)$ is a diagonal matrix contining $\lambda_i$ on its diagonal, and $N$ is nilpotent. 
    Since $\text{Diag}(\lambda_i)$ and $N$ commute, we can write $e^{\Lambda_i t}=e^{\text{Diag}(\lambda_i)t}e^{Nt}$. 
    Since $N$ is nilpotent, every coordinate in $e^{Nt}$ is a polynomial in $t$. 
    Moreover, $e^{\text{Diag}(\lambda_i)t}=\text{Diag}\parenv{e^{\lambda_it}}$. 
    Therefore, if the real part $\Re(\lambda_i)<0$, then as $t\to\infty$ $e^{\Lambda_i t}\to 0$. 

    The only eigenvalues of $\bbM-(\tau+k-1)\bbI$ with non-negative real parts are $0$, 
    which are associated with the left eigenvectors $\bbl_0,\dots \bbl_{s_0-1}$. 
    This means that the only positions that will not decay to $0$ as $t\to\infty$, are the positions in $\Lambda$ that contain $0$, 
    which contain $1$ in $e^{\Lambda}$. 
    This implies that as $t\to\infty$, 
    \[\bbx_t=\E\sparenv{\frac{\bct_n}{n(\tau-1)+m}}\to \sum_{j\in [s_0]}\left\langle \bbl_j,\frac{1}{m}\bct_w\right\rangle \bbr_j\] 
    which concludes the proof. 
\end{IEEEproof}

Our next goal, is to study the frequency of $k$-tuples in cases where $\bbM^{(k)}$ has a unique real eigenvalue $\rho(\bbM^{(k)})$. 

\subsection{Convergence in probability}
%%%%%%%%%%%%%%%%%%%%%%%%%%%%%%%%%%%%%%%%
In this subsection, we focus on the scenario where $\rho(\bbM^{(k)})$ is a simple eigenvalue, indicating that $\bbM^{(k)}$ possesses a unique real eigenvalue with maximum modulus. For instance, Theorem \ref{th:irreducible_mat_properties} specifies that if $\bbM^{(k)}$ is irreducible, then $\rho(\bbM^{(k)})$ is a simple eigenvalue. However, if $\rho(\bbM^{(k)})$ is simple, it does not necessarily imply the irreducibility of $\bbM^{(k)}$. This assumption, together with another technical assumption, facilitates the study of the frequency of $k$-tuples, in contrast to the expected frequency of $k$-tuples explored thus far. In this subsection, we demonstrate that the frequency of $k$-tuples converges in probability to the normalized right eigenvector $\bbr$ of $\bbM^{(k)}$, associated with $\rho(\bbM^{(k)})$. 

Throughout, we assume that $S$ is an average $\tau$-mutation system over the alphabet $\cA=\mathset{a_0,\dots,a_{d-1}}$ of size $d$. We fix $k\in\N$ and denote by $\lambda_0=\tau+k-1$ the unique real eigenvalue of maximum modulus. We also fix a starting word $w$ of length $|w|=m$ with $m\geq k$. 
To simplify notation, we will write $\fr_n(\cdot),\bfr_n,\ct_n(\cdot),\bct_n$ instead of $\fr^{(k)}_{S(n)}(\cdot),\bfr^{(k)}_{S(n)},\ct^{(k)}_{S(n)}(\cdot),\bct^{(k)}_{S(n)}$. 
We also use $\bbM$ instead of $\bbM^{(k)}$ and denote by $Y_n$ the random length of $\vt^n(w)$.

We start by stating the main result of this section. 
For a number $x\in \C$ we denote by $\Re(x)$ the real part of $x$. 
For a matrix $\bbM$ and for an eigenvalue $\lambda$ of $\bbM$, we denote by $\mu_a(\lambda),\mu_g(\lambda)$ the algebraic multiplicity and geometric multiplicity of $\lambda$. 
\begin{theorem} 
\label{th:main3}
Let $S$ be an average $\tau$-mutation system over the alphabet $\cA=\mathset{a_0,\dots,a_{d-1}}$. 
Fix $k$ and let $w$ be a starting word of length $m\geq k$. 
Assume that the $k$-substitution matrix $\bbM^{(k)}$ has a unique real eigenvalues $\lambda_0=\rho(\bbM^{(k)})=\tau+k-1$. 
Let $\bbl$ and $\bbr$ represent the left and right eigenvectors associated with $\lambda_0$, respectively, with normalization such that $\bbr$ forms a probability vector, 
and $\bbl \cdot \bbr = 1$. If $\Re(\lambda)<k$ for every eigenvalue $\lambda$ of $\bbM^{(k)}$ with $\mu_a(\lambda)>\mu_g(\lambda)$, then  
\[\bfr_{S(n)}\stackrel{P}{\longrightarrow} \bbr.\]
\end{theorem}

Before diving into the proof of Theorem \ref{th:main3}, we briefly explain the idea and intuition behind it. Up to this point, we observed that the expected frequency of $k$-tuples evolves according to the $k$-substitution matrix $\bbM$. Drawing insights from urn models, it is reasonable to speculate that, as $\R^{d^k}$ vectors, the frequency $(\bfr_n)_n$ evolves according to the eigenvalues of $\bbM$ in the directions of the eigenvectors of $\bbM$. Thus, if we express $\bfr_n$ with respect to the basis of eigenvectors (and generalized eigenvectors) of $\bbM$, it may be easier to analyze its evolution. It turns out that, except for a single vector, the projection of $\bfr_n$ onto the basis vectors decays to $0$, leaving us with the limiting frequency of $k$-tuples. 
This proof technique first appeared in \cite{smythe1996central} when studying the distribution of balls in generalized urn models. 

The primary effort in this subsection is to demonstrate that for most row vectors $\bbu$, the projection $\bbu\cdot \bfr_n\to 0$. 
It will be more manageable to study the telescopic sum $\sum_{j=1}^n \bbu\cdot (\bfr_j-\bfr_{j-1})$. 
To achieve this, we use Lemma \ref{lem:limlength} to approximate $\bfr_n$ by a $\frac{1}{n(\tau-1)+m}\bct_n$, which is easier to analyze. 
Then we show that the error due to the approximation decays to $0$.

Recall that in Lemma \ref{lem:limlength}, we showed that $Y_n\to n(\tau-1)+m$ almost surely. 
For simplicity, define $\nabla Y_n:=Y_n-Y_{n-1}$ and denote by $\cF_i$ the $\sigma$-algebra generated by $\bct_j$ for $j\leq i$. 
From the proof of Lemma \ref{lem:limlength}, we have that $\E[\nabla Y_i ~|~ \cF_{i-1}]=\tau-1$ and that 
\[\E[\nabla Y_i-\E[\nabla Y_i ~|~ \cF_{i-1}]]=0.\] 

Moreover, since $\cF_i\subseteq \cF_{i+1}$, for $i<j$ we obtain 
\begin{align*}
    \E[(\nabla Y_j-(r-1)) ~|~ \cF_i]&= \E[\E[(\nabla Y_j-(\tau -1)) ~|~ \cF_{j-1}] ~|~ \cF_i]\\ 
    &=0.
\end{align*}
This implies that for $i<j$, 
\begin{align}
    \label{eq:exp_diff_0}
    &\E\sparenv{(\nabla Y_i-\E[\nabla Y_i ~|~ \cF_{i-1}])(\nabla Y_j-\E[\nabla Y_j ~|~ \cF_{j-1}])} \\ \nonumber
    &=0
\end{align}
Thus, we obtain the following useful corollary. 
\begin{corollary}
    \label{cor:bounded_dist}
    Let $Y_n$ denote the length of $S(n)$ and let $\nabla Y_n:=Y_n-Y_{n-1}$. 
    Let $\mathset{\cF_i}_i$ be a filtration as defined above. Then there is a constant $C_0$, such that 
    \[\frac{1}{(n(\tau-1)+m)^2}\E\sparenv{\parenv{Y_n-n\parenv{(\tau-1)+\frac{m}{n}}}^2}\leq \frac{C_0}{n}.\]
\end{corollary}

\begin{IEEEproof}
    Recall that $Y_0=m$ and consider 
    \begin{align}
    \label{eq:cor_proof_1}
        & \E\sparenv{\parenv{\sum_{i=1}^n (\nabla Y_i-(\tau-1))}^2} \\ \nonumber 
        &= \E\sparenv{(Y_n-m-n(\tau-1))^2}\\ \nonumber 
        &= \E\sparenv{\parenv{Y_n-n\parenv{(\tau-1)+\frac{m}{n}}}^2}.
    \end{align} 
    On the other hand, 
\begin{align}
\label{eq:cor_proof_2}
    &\E\sparenv{\parenv{\sum_{i=1}^n (\nabla Y_i-(\tau-1))}^2}\\ \nonumber
    &= \sum_{i=1}^n\E\sparenv{(\nabla Y_i-(\tau-1))^2}+ 2\sum_{1\leq i<j\leq n}\E\sparenv{(\nabla Y_i-(\tau-1))(\nabla Y_j-(\tau-1))}\\ \nonumber
    &\stackrel{(a)}{=} \sum_{i=1}^n\E\sparenv{(\nabla Y_i-(\tau-1))^2}\\ \nonumber 
    &\stackrel{(b)}{\leq} Cn
\end{align} 
where $(a)$ follows from \eqref{eq:exp_diff_0} and $(b)$ follows since $\nabla Y_i-(\tau-1)$ is bounded. 
Plugging \eqref{eq:cor_proof_2} into \eqref{eq:cor_proof_1}, we get 
\begin{align}
\label{eq:lem2}
    &\frac{1}{(n(\tau-1)+m)^2}\E\sparenv{\parenv{ Y_n-n\parenv{(\tau-1)+\frac{m}{n}}}^2}\\ \nonumber 
    &\leq \frac{1}{(n(\tau-1)+m)^2}Cn\\ \nonumber 
    &=\frac{C}{n(\tau-1)^2+O(1)}\\ \nonumber
    &\leq \frac{C_0}{n}
\end{align}
for some constant $C_0$.
\end{IEEEproof}

We will now demonstrate that the only eigenvector influencing the frequency of symbols is the one corresponding to the real eigenvalue $\tau+k-1$. 
Our initial objective is to establish the following claim.
\begin{claim} 
\label{cl:Xn_to_0}
Consider an eigenvalue $\lambda\neq \tau+k-1$, where $\lambda=\lambda_r+\bbi\lambda_c$ with $\lambda_r,\lambda_c\in \R$ representing its real and imaginary parts, respectively. 
Let $\bbu$ be the (normalized) left eigenvector of $\bbM^{(k)}$ corresponding to $\lambda$, expressed as $\bbu=\bbu_r+\bbi \bbu_c$. 
Let $\lambda_0=\tau+k-1$ and define $X_n=X_n^r+iX_n^c$, where $X^r_n=\bbu_r\cdot \bct_{S(n)}$ and $X^c_n=\bbu_c\cdot \bct_{S(n)}$. Then, 
\[\frac{X_n}{n(\tau-1)+m}\stackrel{P}{\to} 0.\]
\end{claim}

\begin{IEEEproof}
    Recall that $Y_j$ denotes the length of $\vt^j(w)$ and $Y_0=|w|=m$. 
    We have 
    \begin{align*}
    \E[X^r_j-X^r_{j-1} ~|~ \cF_{j-1}]&= \E[\bbu_r\cdot \parenv{\bct_j-\bct_{j-1}} ~|~ \cF_{j-1}]\\  
    &= \bbu_r\cdot \E[\parenv{\bct_j-\bct_{j-1}} ~|~ \cF_{j-1}]\\ 
    &\stackrel{(a)}{=} \bbu_r\cdot \frac{1}{ Y_{j-1}}\parenv{\bbM-k\bbI}\cdot\bct_{j-1},
    \end{align*}
    where $(a)$ follows from Lemma \ref{lem:count_of_next_step_mut} and Remark \ref{rem:M_works_in_general}. 
    For a matrix $\bbA$ and an eigenvector $\bbv=\bbv_r+\bbi \bbv_c$ with eigenvalue $\alpha=\alpha_r+\bbi \alpha_c$, we have 
    \[\bbA \bbv_r+\bbi \bbA \bbv_c=\bbA \bbv=\alpha\bbv=(\alpha_r+\bbi \alpha_c)(\bbv_r+\bbi \bbv_c)=(\alpha_r\bbv_r-\alpha_c\bbv_c)+\bbi (\alpha_c\bbv_r+\alpha_r\bbv_c).\] 
    Thus, we obtain 
    \begin{align*} 
    \frac{1}{ Y_{j-1}} \bbu_r\cdot\parenv{\bbM-k\bbI}\cdot\bct_{j-1} 
    &= \frac{\lambda_r-k}{ Y_{j-1}}\bbu_r\cdot\bct_{j-1} - \frac{\lambda_c}{ Y_{j-1}}\bbu_c\cdot\bct_{j-1} \\ 
    &= \frac{\lambda_r-k}{ Y_{j-1}}X_{j-1}^r - \frac{\lambda_c}{ Y_{j-1}}X^c_{j-1}.
\end{align*} 
Similarly, we obtain 
\[\E[X^c_j-X^c_{j-1} ~|~ \cF_{j-1}]=\frac{\lambda_r-k}{ Y_{j-1}}X_{j-1}^c + \frac{\lambda_c}{ Y_{j-1}}X^r_{j-1}.\]

We now define 
\begin{align*}
M_j^r&:=(X^r_j-X^r_{j-1})-\frac{1}{Y_{j-1}}\parenv{(\lambda_r-k)X^r_{j-1}-\lambda_cX^c_{j-1} } \\ 
M_j^c&:=(X^c_j-X^c_{j-1})-\frac{1}{Y_{j-1}}\parenv{(\lambda_r-k)X^c_{j-1}+\lambda_cX^r_{j-1} },
\end{align*}
and notice that $M_j^r,M_j^c$ are martingale differences sequences with $\E[M^r_j]=\E[M^c_j]=0$. 

Our goal is to replace $Y_j$ with $j(\tau-1)+m$ which will be easier to analyze. 
Thus, we define 
\begin{align*}
\overline{M}_j^r&:=(X^r_j-X^r_{j-1})-\frac{1}{(j-1)(\tau-1)+m}\parenv{(\lambda_r-k)X^r_{j-1}-\lambda_cX^c_{j-1} } \\ 
\overline{M}_n^c&:=(X^c_j-X^c_{j-1})-\frac{1}{(j-1)(\tau-1)+m}\parenv{(\lambda_r-k)X^c_{j-1}+\lambda_cX^r_{j-1} }.
\end{align*} 
Of course, $\overline{M}_j^r,\overline{M}_j^c$ may no longer be martingale differences. 
But Lemma \ref{lem:limlength} suggests that such a replacement will not affect any limiting result we obtain. 

We now use $\overline{M}_j^r,\overline{M}_j^c$ to obtain $X_n$. 
We seek constants $\beta^c_{j,n},\beta^r_{j,n}$ such that 
\[V_n:=\sum_{j=1}^n\parenv{\beta^r_{j,n}\overline{M}^r_j+\bbi\beta^c_{j,n}\overline{M}^c_j}=X^r_n+\bbi X^c_n+\varepsilon_n\] 
where $\varepsilon_n$ is a small error term. 

Let $\beta^r_{n,n}=1$ and $\beta^c_{n,n}=1$.
Choose the rest of the coefficients recursively to eliminate $X_{n-1},\dots,X_0$. 
We obtain 
\begin{align*}
    &\beta^r_{n,n}\parenv{-X^r_{n-1}-\frac{1}{(n-1)(\tau-1)+m}\parenv{(\lambda_r-k)X^r_{n-1}-\lambda_cX^c_{n-1} }}\\ 
    &+\bbi\beta^c_{n,n}\parenv{-X^c_{n-1}-\frac{1}{(n-1)(\tau-1)+m}\parenv{(\lambda_r-k)X^c_{n-1}+\lambda_cX^r_{n-1} }}\\\\ 
    &+ \beta^r_{n-1,n}X^r_{n-1}+\bbi\beta^c_{n-1,n}X^c_{n-1}=0,
\end{align*} 
which gives 
\begin{align*}
    \beta_{n-1,n}^r&=\parenv{1+\frac{\lambda_r-k}{(n-1)(\tau-1)+m}}+\bbi\frac{\lambda_c}{(n-1)(\tau-1)+m}\\ 
    \beta_{n-1,n}^c&=\parenv{1+\frac{\lambda_r-k}{(n-1)(\tau-1)+m}}+\bbi\frac{\lambda_c}{(n-1)(\tau-1)+m}.
\end{align*}
In general, for $j\leq n$ we have 
\begin{align}
\label{eq:main_proof_1}
    \beta_{j-1,n}^r&=\parenv{1+\frac{\lambda_r-k}{(j-1)(\tau-1)+m}}\beta^r_{j,n}+\bbi\frac{\lambda_c}{(n-1)(\tau-1)+m}\beta^c_{j,n}\\ \nonumber
    \beta_{j-1,n}^c&=\bbi\frac{\lambda_c}{(n-1)(\tau-1)+m}\beta^r_{j,n}+\parenv{1+\frac{\lambda_r-k}{(j-1)(\tau-1)+m}}\beta^c_{j,n}.
\end{align}

To simplify the analysis of $V_n$, we write the recursion relation in vector form. 
Let $\bbeta_{j,n}=\bmat{\beta^r_{j,n}\\ \beta^c_{j,n}}$ and define $\bbeta_{n,n}=\bmat{1\\ 1}$.
From  \eqref{eq:main_proof_1}, we have  
\begin{align}
    \label{eq:rec1}
    \bbeta_{j-1,n}&= \parenv{\bbI+\frac{1}{(j-1)(\tau-1)+m}\bbA}\bbeta_{j,n} \\ \nonumber 
    &= \prod_{t=j}^n\parenv{\bbI+\frac{1}{(t-1)(\tau-1)+m}\bbA}\bmat{1\\ 1}
\end{align}
where $\bbI$ is a $2\times 2$ identity matrix and 
\[\bbA=\bmat{\lambda_r -k & \bbi\lambda_c \\ \bbi\lambda_c & \lambda_r-k}.\] 

The eigenvalues of $\bbA$ are 
\begin{align*}
    \rho_1&= \lambda_r-k+\bbi |\lambda_c|,\\ 
    \rho_2&= \lambda_r-k-\bbi |\lambda_c|,
\end{align*}
with eigenvectors 
\begin{align*}
    \bbv_1= \bmat{ 1\\ \frac{|\lambda_c|}{\lambda_c}},\; \; \bbv_2= \bmat{1\\ -\frac{|\lambda_c|}{\lambda_c}}.
\end{align*} 
Notice that if $\lambda_c> 0$ then $\rho_1=\lambda-k$ and $\rho_2=\overline{\lambda}-k$, and if $\lambda_c< 0$ then 
$\rho_1=\overline{\lambda}-k$ and $\rho_2=\lambda-k$. 
If $\lambda_c=0$ then $\rho_1=\rho_2=\lambda_r-k$, $\bbA$ is diagonal, and the eigenvalues are the standard basis vectors. 

We distinguish between the three cases. 
\begin{enumerate}
\item Let us assume first that $\lambda_c> 0$. 
Then 
\begin{align*}
    \bbv_1= \bmat{1\\ 1 },\; \; \bbv_2= \bmat{1\\ -1 }.
\end{align*}
If $\bbV$ is the matrix $\bmat{\bbv_1& \bbv_2}$ then we can write 
\[\bbV^{-1}\bbA\bbV=\bmat{\lambda-k & 0 \\ 0 & \overline{\lambda}-k}.\]
Plugging this into \eqref{eq:rec1}, we get  
\begin{align*}
    \bbeta_{j-1,n}&= \bbV \bmat{\prod_{t=j}^n\parenv{1+\frac{\lambda-k}{(t-1)(\tau-1)+m}} & 0 \\ 0& \prod_{t=j}^n\parenv{1+\frac{\overline{\lambda}-k}{(t-1)(\tau-1)+m}}} \bbV^{-1}\bmat{1\\ 1}. 
\end{align*}
Using Stirling's approximation \eqref{eq:stirling_for_gamma_ratio}, we have
\begin{align*}
    \prod_{t=j}^n\parenv{1+\frac{\lambda-k}{(t-1)(\tau-1)+m}}&=\prod_{t=j}^n\parenv{\frac{(t-1)+ 
    \frac{m+\lambda-k}{\tau-1}}{(t-1)+\frac{m}{\tau-1}}} \\ 
    %&=\frac{\Gamma\parenv{n+\frac{m+\lambda-k(1+\bbi)}{\tau-1}}}{\Gamma\parenv{j+\frac{m+\lambda-k(1+\bbi)}{\tau-1}}} \frac{\Gamma\parenv{j+\frac{m}{\tau-1}}}{\Gamma\parenv{n+\frac{m}{\tau-1}}}\\ 
    &= \parenv{\frac{n}{j}}^{(\lambda-k)/(\tau-1)}+O\parenv{n^{\frac{\lambda-k}{\tau-1}-1}}
\end{align*} 
Similarly, we have 
\[\prod_{t=j}^n\parenv{1+\frac{\overline{\lambda}-k}{(t-1)(\tau-1)+m}}=\parenv{\frac{n}{j}}^{\frac{\overline{\lambda}-k}{\tau-1}}+ 
O\parenv{n^{\frac{\overline{\lambda}-k}{\tau-1}-1}}.\]

Writing $\frac{n}{j}$ as $e^{\log (n/j)}$, we get 
\begin{align*}
\parenv{\frac{n}{j}}^{\frac{\lambda-k}{\tau-1}}&= 
\parenv{\frac{n}{j}}^{\frac{\lambda_r-k}{\tau-1}}\parenv{\frac{n}{j}}^{\bbi\frac{\lambda_c}{\tau-1}}\\
&= \parenv{\frac{n}{j}}^{\frac{\lambda_r-k}{\tau-1}}\parenv{\cos\parenv{\frac{\lambda_c}{\tau-1}\log (n/j)} +\bbi\sin\parenv{\frac{\lambda_c}{\tau-1}\log (n/j)}}.
\end{align*}
Similarly, we have  
\[\parenv{\frac{n}{j}}^{\frac{\overline{\lambda}-k}{\tau-1}}=\parenv{\frac{n}{j}}^{\frac{\lambda_r-k}{\tau-1}} \parenv{\cos\parenv{\frac{\lambda_c}{\tau-1}\log (n/j)} -\bbi\sin\parenv{\frac{\lambda_c}{\tau-1}\log (n/j)}}.\]
Plugging this back into $\bbeta_{j-1,n}$ and noticing that $\bbV^{-1}=\frac{1}{2}\bmat{1&1 \\ 1& -1}$ we obtain 
\begin{align*}
    \bbeta_{j-1,n}&= \bmat{
    \parenv{\frac{n}{j}}^{\frac{\lambda_r-k}{\tau-1}} \parenv{\cos\parenv{\frac{\lambda_c}{\tau-1}\log (n/j)} +\bbi\sin\parenv{\frac{\lambda_c}{\tau-1}\log (n/j)}}+O\parenv{n^{\frac{\lambda_r-k}{\tau-1}-1}}\\ 
    \parenv{\frac{n}{j}}^{\frac{\lambda_r-k}{\tau-1}} \parenv{\cos\parenv{\frac{\lambda_c}{\tau-1}\log (n/j)} +\bbi\sin\parenv{\frac{\lambda_c}{\tau-1}\log (n/j)}}+O\parenv{n^{\frac{\lambda_r-k}{\tau-1}-1}}
    }.
\end{align*}

\item Assume now $\lambda_c< 0$ and repeating the exact same calculations, we obtain  
\begin{align*}
    \bbeta_{j-1,n}&= \bmat{
    \parenv{\frac{n}{j}}^{\frac{\lambda_r-k}{\tau-1}} \parenv{\cos\parenv{\frac{\lambda_c}{\tau-1}\log (n/j)} -\bbi\sin\parenv{\frac{\lambda_c}{\tau-1}\log (n/j)}}+O\parenv{n^{\frac{\lambda_r-k}{\tau-1}-1}}\\ 
    \parenv{\frac{n}{j}}^{\frac{\lambda_r-k}{\tau-1}} \parenv{\cos\parenv{\frac{\lambda_c}{\tau-1}\log (n/j)} -\bbi\sin\parenv{\frac{\lambda_c}{\tau-1}\log (n/j)}}+O\parenv{n^{\frac{\lambda_r-k}{\tau-1}-1}}
    }.
\end{align*}

\item Finally, similar calculations for the case $\lambda_c=0$ yields 
\begin{align*}
    \bbeta_{j-1,n}&= \bmat{
    \parenv{\frac{n}{j}}^{\frac{\lambda_r-k}{\tau-1}} +O\parenv{n^{\frac{\lambda_r-k}{\tau-1}-1}}\\ 
    \bbi \parenv{\frac{n}{j}}^{\frac{\lambda_r-k}{\tau-1}} +O\parenv{n^{\frac{\lambda_r-k}{\tau-1}-1}}
    }.
\end{align*}
\end{enumerate}

In all three cases, we have 
\begin{align}
\label{eq:sizebeta}
    \abs{\bbeta_{j-1,n}}&\leq \bmat{\parenv{\frac{n}{j}}^{\frac{\lambda_r-k}{\tau-1}}+O\parenv{n^{\frac{\lambda_r-k}{\tau-1}-1}}\\ 
\parenv{\frac{n}{j}}^{\frac{\lambda_r-k}{\tau-1}}+O\parenv{n^{\frac{\lambda_r-k}{\tau-1}-1}}}.
\end{align}

We now turn to the estimation of $\varepsilon_n$. 
For all three cases $\lambda_c< 0$, $\lambda_c=0$, and $\lambda_c> 0$, we have 
\begin{align}
\label{eq:epsilon_est}
    \varepsilon_n&= -X^r_0\parenv{ \parenv{1+\frac{\lambda_r-k}{m}}\beta^r_{1,n}+\frac{\lambda_c}{m}\bbi\beta^c_{1,n} }- 
    X^c_0\parenv{ \parenv{1+\frac{\lambda_r-k}{m}}\bbi\beta^c_{1,n}-\frac{\lambda_c}{m}\beta^r_{1,n} }\\ \nonumber
    &= O\parenv{n^{\frac{\lambda_r-k}{\tau-1}}}
\end{align}
where the last equality follows since $X_0$ is bounded. 
Since $\lambda_r<\tau+k-1$, we have $\frac{\lambda_r-k}{\tau-1}<1$, which implies $|\varepsilon_n|=o(n)$. 

We now write $V_n$ as the following sum. 
\begin{align*}
    V_n&= \sum_{t=1}^n \parenv{\beta^r_{t,n}M^r_t+\bbi\beta^c_{t,n}M^c_t}\\  
    &\quad +\sum_{t=1}^n\frac{1}{Y_{t-1}}\parenv{\beta^r_{t,n}\parenv{(\lambda_r-k)X^r_{t-1}-\lambda_cX^c_{t-1}}+\bbi\beta^c_{t,n}\parenv{(\lambda_r-k)X^c_{t-1}+\lambda_cX^r_{t-1} }} \\ 
    &\quad -\sum_{t=1}^n\frac{1}{(t-1)(\tau-1)+m}\parenv{\beta^r_{t,n}\parenv{(\lambda_r-k)X^r_{t-1}-\lambda_cX^c_{t-1}}+\bbi\beta^c_{t,n}\parenv{(\lambda_r-k)X^c_{t-1}+ \lambda_cX^r_{t-1} }}.
\end{align*}
Which yields  
\begin{align}
\label{eq:r1}
    |V_n|&\leq \abs{\sum_{t=1}^n \parenv{\beta^r_{t,n}M^r_t+\bbi\beta^c_{t,n}M^c_t}} \\ \nonumber   
    &\quad +\abs{\sum_{t=1}^n\beta^r_{t,n}(\lambda_r-k)X^r_{t-1}\parenv{\frac{1}{Y_{t-1}}-\frac{1}{(t-1)(\tau-1)+m}}} \\ \nonumber  
    &\quad + \abs{\sum_{t=1}^n\beta^r_{t,n}\lambda_cX^c_{t-1}\parenv{\frac{1}{Y_{t-1}}-\frac{1}{(t-1)(\tau-1)+m}}}\\ \nonumber 
    &\quad + \abs{\bbi\sum_{t=1}^n\beta^c_{t,n}(\lambda_r-k)X^c_{t-1}\parenv{\frac{1}{Y_{t-1}}-\frac{1}{(t-1)(\tau-1)+m}}}\\ \nonumber 
    &\quad +\abs{\bbi\sum_{t=1}^n\beta^c_{t,n}\lambda_cX^r_{t-1}\parenv{\frac{1}{Y_{t-1}}-\frac{1}{(t-1)(\tau-1)+m}}}.
\end{align} 

We now show that $\Pr\parenv{\frac{|v_n|}{n(\tau-1)+m}\geq \delta}\to 0$ for any $\delta>0$. 
To that end, we first notice that for $t<j$, 
\begin{align} 
\label{eq:main_proof_2}
    \E\sparenv{\parenv{\beta_t^rM_t^r+\bbi\beta_t^cM_t^c}\parenv{\beta_j^rM_j^r+\bbi\beta_j^cM_j^c}}=0.
\end{align}
This follows from the law of total expectation together with 
\begin{align*}
    \E\sparenv{\beta_t^rM_t^r\beta_j^rM_j^r ~|~ \cF_t}&= \beta_t^rM_t^r\beta_j^r\E\sparenv{M_j^r ~|~ \cF_t}\\ 
    &= \beta_t^rM_t^r\beta_j^r\E\sparenv{\E\sparenv{M_j^r ~|~ \cF_{j-1}} ~|~\cF_t}\\ 
    &=0.
\end{align*} 
Thus,  
\begin{align*}
    \E\sparenv{\beta_t^rM_t^r \bbi\beta_j^cM_j^c}=0, \\
    \E\sparenv{\bbi\beta_t^cM_t^c\beta_j^rM_j^r}=0, \\ 
    \E\sparenv{-\beta_t^cM_t^c\beta_j^cM_j^c}=0.
\end{align*} 
Moreover, since $|M_j^c|, |M_j^r|$ are bounded for every $j$, there exists a constant $C$ such that 
\begin{align}
\label{eq:main_proof_3}
\E\sparenv{\parenv{\beta_j^rM_j^r+\bbi\beta_j^cM_j^c}^2}&\leq C \parenv{\parenv{\frac{n}{j}}^{2\frac{\lambda_r-k}{\tau-1}}+O\parenv{\frac{n^{2\frac{\lambda_r-k}{\tau-1}-1}}{j^{\frac{\lambda_r-k}{\tau-1}}}}+ O\parenv{n^{2\frac{\lambda_r-k}{\tau-1}-2}} }.
\end{align} 
Combining \eqref{eq:main_proof_2} and \eqref{eq:main_proof_3}, we get 
\begin{align*}
    \var\parenv{\sum_{t=1}^n(\beta^r_{t,n}M^r_t+\bbi\beta^c_{t,n}M^c_t)}&= \sum_{t=1}^n \E\sparenv{\parenv{\beta^r_{t,n}M^r_t+\bbi\beta^c_{t,n}M^c_t}^2}\\ 
    &\leq C\sum_{t=1}^n \parenv{\parenv{\frac{n}{t}}^{2\frac{\lambda_r-k}{\tau-1}}+O\parenv{\frac{n^{2\frac{\lambda_r-k}{\tau-1}-1}}{t^{\frac{\lambda_r-k}{\tau-1}}}}+ O\parenv{n^{2\frac{\lambda_r-k}{\tau-1}-2}} }.
\end{align*} 
Since $\lambda\neq \tau+k-1$ (although $|\lambda|$ might be equal to $\tau+k-1$), we have that 
\[2\frac{\lambda_r-k}{\tau-1}<2.\]
Let us denote $2\frac{\lambda_r-k}{\tau-1}=\alpha$, and evaluate the sum $\sum_{t=1}^n \parenv{\frac{n}{t}}^{\alpha}$. 
We get 
\[\sum_{t=1}^n \parenv{\frac{n}{t}}^{\alpha}\approx \begin{cases} 
O\parenv{n^{\alpha}}& \text{ if } \alpha>1\\ 
O\parenv{n\log (n)}& \text{ if } \alpha=1, 
\end{cases}\] 
and 
\[\sum_{t=1}^n \parenv{\frac{n}{t}}^{\alpha}\leq \begin{cases} 
n^{1+\alpha}& \text{ if } 0<\alpha<1\\ 
n& \text{ if } \alpha\leq 0.
\end{cases}\]
Overall, since $2\frac{\lambda_r-k}{\tau-1}<2$, we have that 
\[\sum_{t=1}^n \parenv{\frac{n}{t}}^{2\frac{\lambda_r-k}{\tau-1}}\leq n^{2-\epsilon_1}\] 
for some $\epsilon_1>0$. 
Similar calculations yield 
\[\var\parenv{\sum_{t=1}^n(\beta^r_{t,n}M^r_t+\bbi\beta^c_{t,n}M^c_t)}\leq Cn^{2-\epsilon_1}.\] 
From Chebyshev's inequality (with $\E[M_i^r]=\E[M_i^c]=0$), 
\begin{align*}
\Pr\parenv{\abs{\frac{\sum_{i=1}^n(\beta^r_{i,n}M^r_i+\bbi\beta^c_{i,n}M^c_i)}{n(\tau-1)+m}}> \epsilon}&\leq \frac{Cn^{2-\epsilon_1}}{\epsilon^2 \parenv{n(\tau-1)+m}^2}\\ 
&\to 0.
\end{align*} 
Thus, we bounded above the first argument in \eqref{eq:r1}. 

For the second argument, consider 
\begin{align*}
    &\E\sparenv{\abs{\sum_{t=1}^n\beta^r_{t,n}(\lambda_r-k)X^r_{t-1}\parenv{\frac{1}{Y_{t-1}}-\frac{1}{(t-1)(\tau-1)+m}}}} \\ 
    &\quad \leq \sum_{t=1}^n|\beta^r_{t,n}(\lambda_r-k)|\cdot \E\sparenv{\abs{\frac{X^r_{t-1}}{Y_{t-1}}\parenv{1-\frac{Y_{t-1}}{(t-1)(\tau-1)+m}}}} \\ 
    &\quad \stackrel{(c.s.)}{\leq} \sum_{t=1}^n|\beta^r_{t,n}(\lambda_r-k)|\cdot \parenv{\E\sparenv{\abs{\frac{X^r_{t-1}}{Y_{t-1}}}^2}\E\sparenv{\abs{1-\frac{Y_{t-1}}{(t-1)(\tau-1)+m}}^2}}^{1/2} 
\end{align*} 
where the last inequality is due to Cauchy-Schwartz inequality. 
Notice that 
\[|X_{t-1}^r|^2\leq |X^r_{t-1}+\bbi X^c_{t-1}|^2=|X_{t-1}|^2.\] 
Thus, 
\begin{align*}
&\E\sparenv{\abs{\sum_{t=1}^n\beta^r_{t,n}(\lambda_r-k)X^r_{t-1}\parenv{\frac{1}{Y_{t-1}}-\frac{1}{(t-1)(\tau-1)+m}}}} \\
    &\quad \leq \sum_{t=1}^n|\beta^r_{t,n}(\lambda_r-k)|\cdot \parenv{\E\sparenv{\abs{\frac{X_{t-1}}{Y_{t-1}}}^2}\E\sparenv{\abs{1-\frac{Y_{t-1}}{(t-1)(\tau-1)+m}}^2}}^{1/2}. 
\end{align*}
Since $|X_n|\leq \|\bbu\|_{\infty}Y_n$ and since the eigenvectors are normalized, There exists some constant $D_0$ for which  
\[\E\sparenv{\abs{\frac{X_{t-1}}{Y_{t-1}}}^2}\leq D_0.\] 

Since $Y_t$ is real for every $t$, we have 
\[\abs{1-\frac{Y_{t-1}}{(t-1)(\tau-1)+m}}^2=\parenv{1-\frac{Y_{t-1}}{(t-1)(\tau-1)+m}}^2.\] 
From Corollary \ref{cor:bounded_dist} we have  
\begin{align*}
    \E\sparenv{\parenv{\frac{Y_{t-1}}{(t-1)(\tau-1)+m}-1}^2}&= \E\sparenv{\parenv{\frac{Y_{t-1}-(t-1)(\tau-1)+m}{(t-1)(\tau-1)+m}}^2}\\ 
    &= \frac{1}{((t-1)(\tau-1)+m)^2}\E\sparenv{\parenv{Y_{t-1}-(t-1)\parenv{(\tau-1)+\frac{m}{t-1}}}^2}\\
    &\leq \frac{C_0}{t-1}.
\end{align*}
From \eqref{eq:sizebeta}, there exists some constant $D_1$ such that  
\begin{align}
\label{eq:exp1}
&\E\sparenv{\abs{\sum_{t=1}^n\beta^r_{t,n}(\lambda_r-k)X^r_{t-1}\parenv{\frac{1}{Y_{t-1}}-\frac{1}{(t-1)(\tau-1)+m}}}}\\ \nonumber
&\leq D_1 \sum_{t=1}^n\parenv{\frac{n}{t+1}}^{\frac{\lambda_r-k}{\tau-1}}\frac{1}{(t-1)^{1/2}} +O\parenv{n^{\frac{\lambda_r-k}{\tau-1}}}.
\end{align}
Again, denote $\frac{\lambda_r-k}{\tau-1}=\alpha$ and notice that $\alpha<1$. 
If $\alpha\leq 0$, then 
\begin{align*}
\sum_{t=1}^n\parenv{\frac{n}{t+1}}^{\alpha}\frac{1}{(t-1)^{1/2}}&\leq\sum_{t=1}^n\frac{1}{(t-1)^{1/2}}\\ 
&\leq 2\sqrt{n}\\ 
&=o(n).
\end{align*}
If $0<\alpha<1/2$, then 
\begin{align*}
\sum_{t=1}^n\parenv{\frac{n}{t+1}}^{\alpha}\frac{1}{(t-1)^{1/2}}&\leq \parenv{\frac{n}{2}}^{\alpha}+\sum_{t=1}^n\parenv{\frac{n}{t}}^{\alpha}\frac{1}{t^{1/2}}\\ 
&\leq\parenv{\frac{n}{2}}^{\alpha}+ n^{\alpha}\int_1^n t^{-\alpha-1/2}\mathrm{d}t\\ 
&= \parenv{\frac{n}{2}}^{\alpha}+n^{\alpha}\cdot O\parenv{n^{-\alpha+1/2}}\\ 
&= o(n).
\end{align*} 
If $1/2\leq \alpha\leq 1$ then 
\begin{align*}
\sum_{t=1}^n\parenv{\frac{n}{t+1}}^{\alpha}\frac{1}{(t-1)^{1/2}}&\leq \parenv{\frac{n}{2}}^{\alpha}+\sum_{t=1}^n\parenv{\frac{n}{t}}^{\alpha}\frac{1}{t^{1/2}}\\ 
&\leq\parenv{\frac{n}{2}}^{\alpha}+ n^{\alpha}O(\log (n))\\ 
&= o(n).
\end{align*}
Overall, we have 
\begin{align*}
\E\sparenv{\abs{\sum_{t=1}^n\beta^r_{t,n}(\lambda_r-k)X^r_{t-1}\parenv{\frac{1}{Y_{t-1}}-\frac{1}{(t-1)(\tau-1)+m}}}}=o(n).
\end{align*}
Using Markov's inequality, 
\[\Pr\parenv{\abs{\frac{\sum_{t=1}^n\beta^r_{t,n}(\lambda_r-k)X^r_{t-1}\parenv{\frac{1}{Y_{t-1}}-\frac{1}{(t-1)(\tau-1)+m}}}{n(\tau-1)+m}}>\epsilon}\leq 
\frac{1}{\epsilon n}o(n)\to 0.\]

Similar calculations yield the same result for the rest of the arguments in \eqref{eq:r1}. 
Putting everything together, we obtain that  
\begin{align*}
    \Pr\parenv{\frac{|V_n|}{n(\tau-1)+m}>\epsilon}\to 0. 
\end{align*}

Since $V_n=X_n-\varepsilon_n$, from \eqref{eq:epsilon_est} we immediately get  
\begin{align*}
    \Pr\parenv{\abs{\frac{X_n}{n(\tau-1)+m}}>\epsilon}&\to 0,
\end{align*}
or in other words, 
\[\frac{X_n}{n(\tau-1)+m}\stackrel{P}{\to}0.\]
\end{IEEEproof}

Using the same method as in Claim \ref{cl:Xn_to_0}, we obtain a similar result for generalized eigenvectors. 
\begin{corollary}
    \label{cor:Zn_to_0}
Consider a (not necessarily real) eigenvalue $\lambda\neq r+k-1$ and write $\lambda=\lambda_r+\bbi\lambda_c$. 
Assume that $k>\lambda_r$.  
Consider a left generalize eigenvector $\bbv$ of $\bbM$ associated with $\lambda$, and write $\bbv=\bbv_r+\bbi \bbv_c$.
Define $Z_n=Z_n^r+iZ_n^c$ where $Z^r_n=\bbv_r\cdot \bct_{\vt^n(w)}$ and $Z^c_n=\bbv_c\cdot \bct_{\vt^n(w)}$. 
Then 
\[\frac{Z_n}{n(\tau-1)+m}\stackrel{P}{\to} 0.\]
\end{corollary}

\begin{IEEEproof} 
    The proof is similar to the proof of Claim \ref{cl:Xn_to_0}. 
    Throughout, we keep the notations from the proof of Claim \ref{cl:Xn_to_0}. 
    Specifically, we use $\ct_j,\bct_j,\fr_j,\bfr_j$, we let $\bbu$ be a normalized left \underline{eigenvector} of $\bbM$ that corresponds to $\lambda$, and we let $X_n=\bbu\cdot\bct_n$. 

    Assume first that $\bbv$ is a first order generalized eigenvector, so $\bbv\cdot \bbM=\lambda\bbv+\alpha\bbu$ where $\bbv$ can be normalized such that $\alpha=1$. 
    We have 
    \begin{align*}
    \E[Z^r_j-Z^r_{j-1} ~|~ \cF_{j-1}]&= \E[\bbv_r\cdot \parenv{\bct_j-\bct_{j-1}} ~|~ \cF_{j-1}]\\  
    &= \bbv_r\cdot \E[\parenv{\bct_j-\bct_{j-1}} ~|~ \cF_{j-1}]\\ 
    &= \frac{1}{ Y_{j-1}} \bbv_r\cdot \parenv{\bbM-k\bbI}\cdot\bct_{j-1}.
    \end{align*} 
    Thus,  
    \begin{align*} 
    \frac{1}{ Y_{j-1}} \bbv_r\cdot\parenv{\bbM -k\bbI}\cdot\bct_{j-1} 
    &= \frac{(\lambda_r-k) \bbv_r-\lambda_c \bbv_c}{ Y_{j-1}}\cdot\bct_{j-1} + \frac{1}{ Y_{j-1}}\bbu_r\cdot\bct_{j-1} \\ 
    &= \frac{(\lambda_r-k)Z_{j-1}^r-\lambda_cZ^c_{j-1}}{ Y_{j-1}} + \frac{1}{ Y_{j-1}}X^r_{j-1}.
\end{align*} 
Similarly, we obtain 
\[\E[Z^c_j-Z^c_{j-1} ~|~ \cF_{j-1}]=\frac{(\lambda_r-k)Z_{j-1}^c+\lambda_c Z^r_{j-1}}{Y_{j-1}} + \frac{1}{ Y_{j-1}}X^c_{j-1}.\] 

Define 
\begin{align*}
N_j^r&:=(Z^r_j-Z^r_{j-1})-\frac{1}{Y_{j-1}}\parenv{(\lambda_r-k)Z_{j-1}^r-\lambda_cZ^c_{j-1}+X^r_{j-1}} \\ 
N_j^c&:=(Z^c_j-Z^c_{j-1})-\frac{1}{Y_{j-1}}\parenv{(\lambda_r-k)Z_{j-1}^c+\lambda_c Z^r_{j-1}+X^c_{j-1}},
\end{align*}
and notice that $N_j^r,N_j^c$ are martingale differences sequences with $\E[N^r_j]=\E[N^c_j]=0$. 
Define $\overline{N}_j^r,\overline{N}_j^c$ by 
\begin{align*}
\overline{N}_j^r&:=(Z^r_j-Z^r_{j-1})-\frac{1}{Y_{j-1}}\parenv{(\lambda_r-k)Z_{j-1}^r-\lambda_cZ^c_{j-1}} \\ 
\overline{N}_j^c&:=(Z^c_j-Z^c_{j-1})-\frac{1}{Y_{j-1}}\parenv{(\lambda_r-k)Z_{j-1}^c+\lambda_c Z^r_{j-1}}.
\end{align*} 
Notice that $X_j^r,X_j^c$ are removed.

We seek constants $\beta^c_{j,n},\beta^r_{j,n}$ such that 
\begin{align}
\label{eq:v_n}
V_n&:=\sum_{j=1}^n\parenv{\beta^r_{j,n}\overline{N}^r_j+\bbi\beta^c_{j,n}\overline{N}^c_j}=Z^r_n+\bbi Z^c_n+\varepsilon_n
\end{align}
where $\varepsilon_n$ is a small error term.  
Using the exact same $\beta^r_{j,n},\beta^c_{j,n}$ as in the proof of Claim \ref{cl:Xn_to_0}, we obtain \eqref{eq:v_n} with 
\begin{align}
    \varepsilon_n&= -Z^r_0\parenv{ \parenv{1+\frac{\lambda_r-k}{m}}\beta^r_{1,n}+\frac{\lambda_c}{m}\bbi\beta^c_{1,n} }- 
    Z^c_0\parenv{ \parenv{1+\frac{\lambda_r-k}{m}}\bbi\beta^c_{1,n}-\frac{\lambda_c}{m}\beta^r_{1,n} }\\ \nonumber
    &= O\parenv{n^{\frac{\lambda_r-k}{\tau-1}}}.
\end{align}
We write $V_n$ as 
\begin{align*}
    V_n&= \sum_{j=1}^n \parenv{\beta^r_{j,n}N^r_j+\bbi\beta^c_{j,n}N^c_j}\\  
    &\quad +\sum_{j=1}^n\frac{1}{Y_{j-1}}\parenv{\beta^r_{j,n}\parenv{(\lambda_r-k)Z^r_{j-1}-\lambda_cZ^c_{j-1}}+\bbi\beta^c_{j,n}\parenv{(\lambda_r-k)Z^c_{j-1}+\lambda_cZ^r_{j-1} }} \\ 
    &\quad -\sum_{j=1}^n\frac{1}{(j-1)(\tau-1)+m}\parenv{\beta^r_{j,n}\parenv{(\lambda_r-k)Z^r_{j-1}-\lambda_cZ^c_{j-1}}+\bbi\beta^c_{j,n}\parenv{(\lambda_r-k)Z^c_{j-1}+ \lambda_cZ^r_{j-1} }}\\ 
    &\quad +f(X_0,X_1,\dots,X_{n-1}),
\end{align*} 
where 
\begin{align*}
f(X_0,X_1,\dots,X_{n-1})&= \sum_{j=0}^{n-1}\frac{1}{Y_j}\parenv{X_j^r\beta^r_{j,n}+ \bbi X_j^c\beta^c_{j,n}}.
\end{align*}

As evident from \eqref{eq:r1}, to demonstrate that $\Pr\parenv{\frac{V_n}{n(\tau-1)+m}\geq\delta}\to 0$ for every $\delta>0$, it is sufficient to establish that $f(X_0,X_1,\dots,X_{n-1})=o(n)$. 
The boundedness of $\abs{\frac{X_j}{Y_j}}$, coupled with the estimates of $\beta_j^r$ and $\beta_j^c$ and the assumption that $k>\lambda_r$ (implying $\frac{\lambda_r-k}{\tau-1}<0$), collectively show that $f(X_0,X_1,\dots,X_{n-1})=o(n)$.

To conclude the proof, a similar strategy is employed to show that $\frac{1}{n(\tau-1)+m}Z_n\stackrel{P}{\rightarrow}0$ for any ordered generalized eigenvector. 
This is accomplished by substituting $\bbu$ with a generalized eigenvector $\bbu'$, leveraging the boundedness of $\abs{\frac{\bbu'\cdot\bct_j}{Y_j}}$, and employing the same estimations for $\beta_j^r$ and $\beta_j^c$.
\end{IEEEproof}
\vspace{0.3cm}
We are now ready to prove the main result of this section. 
\begin{IEEEproof}[Proof of Theorem \ref{th:main3}] 
    Let $\bbl,\bbr$ be the left and right eigenvectors associated with $\rho(\bbM^{(k)})$, respectively. 
    Recall that $\bbl=\1$ is the vector of all ones. 
    Let $\bbu_j$ for $j=1,\dots,d^k-1$ be the left eigenvectors and generalized eigenvectors obtained by the Jordan decomposition of $\bbM^{(k)}$. 

    The vectors $\bbl$ and $\mathset{\bbu_j}_{j=1}^{d^k-1}$ form a basis for $\R^{d^k}$. Writing a vector $\eta$ with respect to this basis, we obtain 
    \[\eta= \alpha \bbl+ \sum_{j=1}^{d^k-1} \alpha_j\bbu_j.\] 
    Since the vector $\bbr$ is orthogonal to $\bbu_j$ for $j=1,\dots,d^k-1$, and since $\bbl\cdot\bbr=1$, we get 
    \[\eta\cdot \bbr=\alpha.\] 
    We write 
    \begin{align*}
        \frac{1}{n(\tau-1)+m}\eta \cdot \bct_{S(n)} &= \frac{1}{n(\tau-1)+m}\alpha \bbl \cdot \bct_{S(n)}+ \frac{1}{n(\tau-1)+m}\sum_{j=1}^{d^k-1}\alpha_j \bbu_j \cdot \bct_{S(n)}\\ 
        &= \frac{1}{n(\tau-1)+m}\alpha \|\bct_{S(n)}\|_1 + \frac{1}{n(\tau-1)+m}\sum_{j=1}^{d^k-1}\alpha_j \bbu_j \cdot\bct_{S(n)}.
    \end{align*}  
    For $j\in\mathset{1,2,\dots,d^k-1}$, $\bbu_j$ is either an eigenvector or a generalized eigenvector. 
    Thus, $\bbu_j\cdot \bct_{S(n)}$ is either $X_n$ or $Z_n$, as defined in Claim \ref{cl:Xn_to_0} and Corollary \ref{cor:Zn_to_0}. 
    This implies that 
    \[\sum_{j=1}^{d^k-1}\alpha_j \frac{\bbu_j \cdot\bct_{S(n)}}{n(\tau-1)+m}\stackrel{P}{\longrightarrow} 0\] 
    as $n\to\infty$. 
    From Lemma \ref{lem:limlength}, $\|\bct_{S(n)}\|_1=n(\tau-1)+m$ almost surely, which implies  
    \[\frac{1}{n(\tau-1)+m}\eta \cdot \bct_{S(n)}\stackrel{P}{\longrightarrow} \alpha=\eta\cdot\bbr.\]
    The result is obtained by taking $\eta$ to be the standard basis vectors. 
\end{IEEEproof}

\section{Conclusion}\label{sec:conc}
%%%%%%%%%%%%%%%%%%%%%%%%%%%%%%%%%%%%%%%%%%%%%%
In this paper, we delved into the analysis of the frequency of $k$-tuples in mutation systems, motivated by in-vivo DNA storage systems. 
The mutation systems we defined represent a specific sub-family of mutations found in real DNA, encompassing duplication systems and duplication with substitution systems.

Our primary focus was on estimating the expected frequency of $k$-tuples in mutation systems, particularly when imposing restrictions on the length of mutations. 
Under such constraints, we successfully derived explicit expressions for the expected frequency of $k$-tuples at each step. 
When transitioning to a more relaxed restriction in the form of an average length, we identified the limiting expected frequency of $k$-tuples. 
In a scenario where the substitution matrix exhibited a unique real maximal eigenvalue and satisfied an additional technical condition, we demonstrated the convergence in probability of the $k$-tuples frequency. Across all instances, we observed a clear connection between the $k$-tuples frequency and the eigenvectors associated with maximal real eigenvalues of the substitution matrix.

As a direction for future research, delving into the study of general mutation systems without specific restrictions on mutation length would be intriguing. 
The convergence of $k$-tuple frequencies in the general case remains an open question. 
Additionally, we believe that it is possible to eliminate the technical restriction inherent in the analysis of average $\tau$-mutation systems, 
aiming to establish the generality of our results for average $\tau$-mutation systems.

\appendix 
%%%%%%%%%%%%%%%

\begin{IEEEproof}[Proof of Lemma \ref{lem:count_of_next_step_mut}] 
    While the result may be intuitive using the definition of $\bbM^{(k)}$, 
    we provide a rigorous proof. 
    We start with explicitly writing the expected value. 
For simplicity, we denote the length of $w$ by $|w|=m$. 
    \begin{align}
    \E\sparenv{\ct_{\vt(w)}(u)}&= \sum_{\eta\in\cAS}\Pr(\vt(w)=\eta)\ct_{\eta}(u) \nonumber \\ 
    %&= \sum_{\eta\in\cA^k}\sum_{t\in [d]}\sum_{i=0}^{m-1}\frac{1}{m}\Pr(\vt(w_i)=\eta)\bo_{[w_i=a_t]}\ct_{w_0^{i-1}\eta w_{i+1}^{m-1}}(u)\\ 
    &= \sum_{l\in\N}\sum_{\eta\in\cA^l}\sum_{i=0}^{m-1}\sum_{x,y\in \cA^k}\frac{1}{m}\Pr(\vt(w_i)=\eta)\bo_{[w_{i-k+1}^i=x]}\bo_{[w_i^{i+k-1}=y]} 
    \ct_{w_0^{i-1}\eta w_{i+1}^{m-1}}(u). \nonumber 
    \end{align}
    Taking coordinates modulo $n$, we may split the sum into two sums, according to whether the $k$-tuple we count overlaps $\eta$ or not, to get 
    \begin{align}
    &\E\sparenv{\ct_{\vt(w)}(u)} \\ 
    &= \sum_{l\in\N}\sum_{\eta\in\cA^l}\sum_{i=0}^{m-1}\sum_{j=0}^{m-k-1}\frac{1}{m}\Pr(\vt(w_i)=\eta) \bo_{[(w_{i+1}^{m-1}w_0^{i-1})_{j+[k]}=u]} \label{eq:4}\\ 
    &\quad + \sum_{l\in\N}\sum_{\eta\in\cA^l}\sum_{x,y\in\cA^k}\sum_{i=0}^{m-1}\sum_{j=0}^{k-2+l}\frac{1}{m}\Pr(\vt(w_i)=\eta)\bo_{[w_{i-k+1}^i=x]}\bo_{[w_i^{i+k-1}=y]} 
    \bo_{[(x_0^{k-2}\eta y_1^{k-1})_{j+[k]}=u]} \label{eq:5}
\end{align}

We can write (\ref{eq:5}) as follows. 
\begin{align*}
    &\sum_{l\in\N}\sum_{\eta\in\cA^l}\sum_{x,y\in\cA^k}\sum_{i=0}^{m-1}\sum_{j=0}^{k-2+l}\frac{1}{m}\Pr(\vt(w_i)=\eta)\bo_{[w_{i-k+1}^i=x]}\bo_{[w_i^{i+k-1}=y]} 
    \bo_{[(x_0^{k-2}\eta y_1^{k-1})_{j+[k]}=u]}\\ 
    &= \sum_{l\in\N}\sum_{\eta\in\cA^l}\sum_{x,y\in\cA^k}\sum_{i=0}^{m-1}\sum_{t\in [d]}\frac{1}{m}\Pr(\vt(a_t)=\eta)\bo_{[w_{i-k+1}^i=x]}\bo_{[w_i^{i+k-1}=y]} \bo_{[w_i=a_t]} 
    \sum_{j=0}^{k-2+l}\bo_{[(x_0^{k-2}\eta y_1^{k-1})_{j+[k]}=u]}. 
    \end{align*}
    We split the sum over $j$ to obtain 
    \begin{align}
    &\sum_{l\in\N}\sum_{\eta\in\cA^l}\sum_{x,y\in\cA^k}\sum_{i=0}^{m-1}\sum_{t\in [d]}\frac{1}{m}\Pr(\vt(a_t)=\eta)\bo_{[w_{i-k+1}^i=x]}\bo_{[w_i^{i+k-1}=y]} \bo_{[w_i=a_t]} 
    \sum_{j=0}^{l-1}\bo_{[(x_0^{k-2}\eta y_1^{k-1})_{j+[k]}=u]} \nonumber \\ 
    &\quad +\sum_{l\in\N}\sum_{\eta\in\cA^l}\sum_{x,y\in\cA^k}\sum_{i=0}^{m-1}\sum_{t\in [d]}\frac{1}{m}\Pr(\vt(a_t)=\eta)\bo_{[w_{i-k+1}^i=x]}\bo_{[w_i^{i+k-1}=y]} \bo_{[w_i=a_t]} 
    \sum_{j=l}^{k-2}\bo_{[(x_0^{k-2}\eta y_1^{k-1})_{j+[k]}=u]} \nonumber \\ 
    &\quad + \sum_{l\in\N}\sum_{\eta\in\cA^l}\sum_{x,y\in\cA^k}\sum_{i=0}^{m-1}\sum_{t\in [d]}\frac{1}{m}\Pr(\vt(a_t)=\eta)\bo_{[w_{i-k+1}^i=x]}\bo_{[w_i^{i+k-1}=y]} \bo_{[w_i=a_t]} \sum_{j=k-1}^{k+l-2}\bo_{[(x_0^{k-2}\eta y_1^{k-1})_{j+[k]}=u]}\nonumber \\ 
    &= \sum_{l\in\N}\sum_{\eta\in\cA^l}\sum_{x\in\cA^k}\sum_{i=0}^{m-1}\sum_{t\in [d]}\frac{1}{m}\Pr(\vt(a_t)=\eta)\bo_{[w_{i-k+1}^i=x]} \bo_{[w_i=a_t]} 
    \sum_{j=0}^{l-1}\bo_{[(x_0^{k-2}\eta)_{j+[k]}=u]} \label{eq:6}\\ 
    &\quad +\sum_{l\in\N}\sum_{\eta\in\cA^l}\sum_{x,y\in\cA^k}\sum_{i=0}^{m-1}\sum_{t\in [d]}\frac{1}{m}\Pr(\vt(a_t)=\eta)\bo_{[w_{i-k+1}^i=x]}\bo_{[w_i^{i+k-1}=y]} \bo_{[w_i=a_t]} 
    \sum_{j=l}^{k-2}\bo_{[x_j^{k-2}\eta y_1^{j-l+1}=u]} \label{eq:7} \\ 
    &\quad + \sum_{l\in\N}\sum_{\eta\in\cA^l}\sum_{y\in\cA^k}\sum_{i=0}^{m-1}\sum_{t\in [d]}\frac{1}{m}\Pr(\vt(a_t)=\eta)\bo_{[w_i^{i+k-1}=y]} \bo_{[w_i=a_t]} 
    \sum_{j=0}^{l-1}\bo_{[(\eta y_1^{k-1})_{j+[k]}=u]}. \label{eq:8}  
\end{align}

We first focus on \eqref{eq:6} and \eqref{eq:8} since those are simpler. 
Considering \eqref{eq:6}, we have 
\begin{align*}
    &\sum_{l\in\N}\sum_{\eta\in\cA^l}\sum_{x\in\cA^k}\sum_{i=0}^{m-1}\sum_{t\in [d]}\frac{1}{m}\Pr(\vt(a_t)=\eta)\bo_{[w_{i-k+1}^i=x]} \bo_{[w_i=a_t]} 
    \sum_{j=0}^{l-1}\bo_{[(x_0^{k-2}\eta)_{j+[k]}=u]} \\ 
    &= \sum_{l\in\N}\sum_{\eta\in\cA^l}\sum_{t\in [d]}\sum_{j=0}^{l-1}\sum_{x\in\cA^k}\frac{1}{m}\Pr(\vt(a_t)=\eta) \bo_{[x_{k-1-j}=a_t]} 
    \ct_w(x)\bo_{[x_0^{k-j-2}\eta_0^j=u]}\\ 
    &= \sum_{l\in\N}\sum_{\eta\in\cA^l}\sum_{t\in [d]}\sum_{j=k-l}^{k-1}\sum_{x\in\cA^k}\frac{1}{m}\Pr(\vt(a_t)=\eta) \bo_{[x_j=a_t]} 
    \bo_{[x_0^{j-1}\eta_0^{k-j-1}=u]}\ct_w(x).
\end{align*}

Considering \eqref{eq:8}, we obtain. 
\begin{align*}
    &\sum_{l\in\N}\sum_{\eta\in\cA^l}\sum_{y\in\cA^k}\sum_{i=0}^{m-1}\sum_{t\in [d]}\frac{1}{m}\Pr(\vt(a_t)=\eta)\bo_{[w_i^{i+k-1}=y]} \bo_{[w_i=a_t]} 
    \sum_{j=0}^{l-1}\bo_{[(\eta y_1^{k-1})_{j+[k]}=u]} \\ 
    &= \sum_{l\in\N}\sum_{\eta\in\cA^l}\sum_{t\in [d]}\sum_{x\in\cA^k}\sum_{j=0}^{l-1}\frac{1}{m}\Pr(\vt(a_t)=\eta)\bo_{[x_0=a_t]} \ct_w(x)
    \bo_{[\eta_j^{l-1} x_1^{k-l+j}=u]}
\end{align*}

We now turn to \eqref{eq:7} and recall that the sum over $j$ is not empty only when $l\leq k$. 
We have 
\begin{align*}
    &\sum_{l\in\N}\sum_{\eta\in\cA^l}\sum_{x,y\in\cA^k}\sum_{i=0}^{m-1}\sum_{t\in [d]}\frac{1}{m}\Pr(\vt(a_t)=\eta)\bo_{[w_{i-k+1}^i=x]}\bo_{[w_i^{i+k-1}=y]} \bo_{[w_i=a_t]} 
    \sum_{j=l}^{k-2}\bo_{[x_j^{k-2}\eta y_1^{j-l+1}=u]} \\ 
    &= \sum_{l\in\N}\sum_{\eta\in\cA^l}\sum_{x,y\in\cA^k}\sum_{i=0}^{m-1}\sum_{t\in [d]}\sum_{j=l}^{k-2}\sum_{z\in\cA^{k-l+1}}\frac{1}{m}\Pr(\vt(a_t)=\eta)\bo_{[w_{i-k+1}^i=x]}\bo_{[w_i^{i+k-1}=y]} \bo_{[w_i=a_t]}\bo_{[x_j^{k-2}\eta y_1^{j-l+1}=u]}\sum_{z\in\cA^{k-l+1}}\bo_{[w_{i+j-k+1}^{i+j-l+1}=z]}.
\end{align*} 
In the above expression, $w_i=z_{k-j-1}$ and $x_j^{k-2}=z_0^{k-j-2}$. Thus, the expression becomes 
\begin{align*}
    &\sum_{l\in\N}\sum_{\eta\in\cA^l}\sum_{x,y\in\cA^k}\sum_{i=0}^{m-1}\sum_{t\in [d]}\sum_{j=l}^{k-2}\sum_{z\in\cA^{k-l+1}}\frac{1}{m}\Pr(\vt(a_t)=\eta)\bo_{[w_{i-k+1}^i=x]}\bo_{[w_i^{i+k-1}=y]} \bo_{[z_{k-j-1}=a_t]}\bo_{[w_{i+j-k+1}^{i+j-l+1}=z]}\bo_{[z_0^{k-j-2}\eta z_{k-j}^{k-l}=u]} \\ 
    &= \sum_{l\in\N}\sum_{\eta\in\cA^l}\sum_{i=0}^{m-1}\sum_{t\in [d]}\sum_{j=l}^{k-2}\sum_{z\in\cA^{k-l+1}}\frac{1}{m}\Pr(\vt(a_t)=\eta) \bo_{[z_{k-j-1}=a_t]}
    \bo_{[w_{i+j-k+1}^{i+j-l+1}=z]}\bo_{[z_0^{k-j-2}\eta z_{k-j}^{k-l}=u]}. 
\end{align*}
Adjusting the sum over $j$ to start with $j=1$, we get 
\[\sum_{l\in\N}\sum_{\eta\in\cA^l}\sum_{i=0}^{m-1}\sum_{t\in [d]}\sum_{j=1}^{k-l-1}\sum_{z\in\cA^{k-l+1}}\frac{1}{m}\Pr(\vt(a_t)=\eta) \bo_{[z_j=a_t]}
    \bo_{[w_{i-j}^{i+k-l-j}=z]}\bo_{[z_0^{j-1}\eta z_{j+1}^{k-l}=u]}.\]
Writing everything with respect to $z$, we obtain 
\begin{align}
    &\sum_{l\in\N}\sum_{\eta\in\cA^l}\sum_{i=0}^{m-1}\sum_{t\in [d]}\sum_{j=1}^{k-l-1}\sum_{z\in\cA^{k-l+1}}\sum_{z'\in\cA^{l-1}}\frac{1}{m}\Pr(\vt(a_t)=\eta) \bo_{[z_j=a_t]}
    \bo_{[w_{i-j}^{i+k-l-j}=z]}\bo_{[w_{i+k-l-j+1}^{i+k-j-1}=z']}\bo_{[z_0^{j-1}\eta z_{j+1}^{k-l}=u]}\nonumber \\ 
    &= \sum_{l\in\N}\sum_{\eta\in\cA^l}\sum_{i=0}^{m-1}\sum_{t\in [d]}\sum_{j=1}^{k-l-1}\sum_{z\in\cA^k}\frac{1}{m}\Pr(\vt(a_t)=\eta) \bo_{[z_j=a_t]}
    \bo_{[w_{i-j}^{i-j+k-1}=z]}\bo_{[z_0^{j-1}\eta z_{j+1}^{k-l}=u]}. \label{eq:9}
\end{align}
Notice that 
\begin{align*}
    \sum_{i=0}^{m-1}\bo_{[w_{i-j}^{i-j+k-1}=z]}&=\ct_w (z), 
\end{align*}
so \eqref{eq:9} becomes 
\begin{align*}
    \sum_{l\in\N}\sum_{\eta\in\cA^l}\sum_{t\in [d]}\sum_{j=1}^{k-l-1}\sum_{z\in\cA^k}\frac{1}{m}\Pr(\vt(a_t)=\eta) \bo_{[z_j=a_t]}
    \bo_{[z_0^{j-1}\eta z_{j+1}^{k-l}=u]}\ct_w(z).
\end{align*}
Putting everything together, we get 
\begin{align*}
    \E\sparenv{\ct_{\vt(w)}(u)}&= \sum_{l\in\N}\sum_{\eta\in\cA^l}\sum_{i=0}^{m-1}\sum_{j=0}^{m-k-1}\frac{1}{m}\Pr(\vt(w_i)=\eta) \bo_{[(w_{i+1}^{m-1}w_0^{i-1})_{j+[k]}=u]}\\ 
    &+ \sum_{l\in\N}\sum_{\eta\in\cA^l}\sum_{t\in [d]}\sum_{x\in\cA^k}\frac{1}{m}\Pr(\vt(a_t)=\eta)\parenv{\sum_{j=1}^{k-1}\bo_{[x_j=a_t]}\bo_{[(x_0^{j-1}\eta x_{j+1}^{k-1})_{[k]}=u]} +\sum_{j=0}^{l-1}\bo_{[x_0=a_t]}\bo_{[\eta_j^{l-1} x_1^{k-l+j}=u]}}\ct_w(x).
\end{align*}
The result follows by noticing that 
\[\sum_{l\in\N}\sum_{\eta\in\cA^l}\sum_{i=0}^{m-1}\sum_{j=0}^{m-k-1}\Pr(\vt(w_i)=\eta) \bo_{[(w_{i+1}^{m-1}w_0^{i-1})_{j+[k]}=u]}=(m-k)\ct_w(u).\]
\end{IEEEproof}

\bibliographystyle{IEEEtranS}
\bibliography{allbib.bib}
\end{document}